\newif\ifTAS
\TASfalse 

\ifTAS
\documentclass[12pt]{article}
\else
\documentclass[11pt,a4paper]{article}
\fi

\usepackage[utf8]{inputenc}
\ifTAS
\else
\usepackage[margin=1in]{geometry}
\usepackage{authblk}
\fi
\usepackage[style=apa]{biblatex}
\usepackage[american]{babel}
\usepackage{listings}
\usepackage{xcolor}  
\definecolor{juliagreen}{rgb}{0,0.5,0}
\definecolor{juliastring}{rgb}{0.6,0.1,0.1}
\definecolor{juliacomment}{gray}{0.4}

\lstdefinelanguage{Julia}{
	morekeywords={
		using,import,function,for,while,if,else,elseif,return,end,begin,
		struct,mutable,let,in,do,try,catch,finally,break,continue,global,local,const
	},
	sensitive=true,
	morecomment=[l]\#,
	morestring=[b]",
}

\usepackage{csquotes}
\DeclareLanguageMapping{american}{american-apa}
\bibliography{references.bib}

\usepackage{threeparttable}
\usepackage{multirow}
\usepackage{multicol}
\usepackage{booktabs}

\usepackage{pifont}
\usepackage{amsmath,amsfonts,amssymb,bm,bbm,amsthm}
\usepackage{graphicx}
\usepackage{verbatim}
\usepackage{nicefrac}

\theoremstyle{definition} 
\newtheorem{prop}{Proposition}

\newtheoremstyle{case}{}{}{}{}{}{:}{ }{}
\theoremstyle{case}

\usepackage[colorlinks=true, allcolors=blue, bookmarks=false]{hyperref}

\usepackage{chngcntr}
\counterwithout{table}{section}
\ifTAS
\newenvironment{revision}{\color{black}}{\color{black}}
\newcommand{\revise}{\textcolor{black}}
\newcommand{\reviseAS}{\textcolor{black}}

\newenvironment{revisionTwo}{\color{blue}}{\color{black}}
\newcommand{\reviseAStwo}{\textcolor{blue}}

\else
\newenvironment{revision}{\color{black}}{\color{black}}
\newcommand{\revise}{\textcolor{black}}
\newcommand{\reviseAS}{\textcolor{black}}

\newenvironment{revisionTwo}{\color{black}}{\color{black}}
\newcommand{\reviseAStwo}{\textcolor{black}}
\fi

\newcommand{\xmark}{\ding{55}}

\newcommand{\prior}[1]{\pi\left({#1}\right)}

\newcommand{\FBeta}[2]{\text{B}\left({#1},\ {#2}\right)}

\newcommand{\BetaBinom}[4]{\text{BB}\left(#1 \mid #2 ,\ #3 ,\ #4 \right)}
\DeclareRobustCommand{\stirlingone}{\genfrac[]{0pt}{}}

\DeclareRobustCommand{\stirling}{\genfrac\{\}{0pt}{}}

\DeclareRobustCommand{\stirling}{\genfrac\{\}{0pt}{}}
\newcommand{\rstirling}[3]{\stirling{#1}{#2}_{#3}}
\newcommand{\bellnum}[1]{B_{#1}}
\newcommand{\rbellnum}[2]{B_{#1,\,#2}}
\newcommand{\setsize}[1]{|{#1}|}

\newcommand{\len}{r} 
\newcommand{\lenj}{{r_{\reviseAStwo{j}}}} 

\newcommand{\partition}{\rho}
\newcommand{\Rho}{\mathrm{P}} 

\newcommand{\shortimplies}{\Rightarrow}

\newcommand{\numberthis}{\addtocounter{equation}{1}\tag{\theequation}}

\ifTAS%
\else

\fi

\ifTAS
\date{}
\else
\title{Flexible Bayesian Multiple Comparison Adjustment Using Dirichlet Process and Beta-Binomial Model Priors}

\author[1]{Don van den Bergh}
\author[2,3]{Fabian Dablander}
\affil[1]{Department of Psychological Methods, University of Amsterdam}
\affil[2]{Institute for Biodiversity and Ecosystem Dynamics, University of Amsterdam}
\affil[3]{Institute for Advanced Study, University of Amsterdam}
\date{\today}
\fi

\ifTAS
\newcommand{\refAppCode}{D}
\newcommand{\refAppDecrDpp}{A.1}
\newcommand{\refAppDecrBb}{A.2}
\newcommand{\refAppD}{D}
\newcommand{\refAppExtraSim}{C}
\newcommand{\refAppMpm}{D}
\newcommand{\refAppStdev}{F}
\newcommand{\refPredictionRuleProof}{A.3}

\newcommand{\refSecProps}{5.1}
\newcommand{\refSimStudy}{4}
\newcommand{\refAppications}{5}

\newcommand{\refDirichletPrior}{7}

\else
\newcommand{\refAppCode}{\ref{sec:appendix-code}}
\newcommand{\refAppDecrDpp}{\ref{ap:decreasing-odds-dpp}}
\newcommand{\refAppDecrBb}{\ref{ap:decreasing-odds}}
\newcommand{\refAppD}{\ref{ap:py_vs_dpp}}
\newcommand{\refAppExtraSim}{\ref{app:simulation}}
\newcommand{\refAppMpm}{\ref{ap:single_model}}
\newcommand{\refAppStdev}{\ref{app:standard-deviation}}
\newcommand{\refPredictionRuleProof}{\label{app:prediction-rule-proofs}}
\newcommand{\refSecProps}{\ref{sec:app:proportions}}
\newcommand{\refSimStudy}{\ref{sec:simulation-study}}
\newcommand{\refAppications}{\ref{sec:applications}}

\newcommand{\refDirichletPrior}{\ref{eq:prediction-rule-DP}}

\fi

\newcommand{\coluniform}{gray}

\newcommand{\colbbonek}{blue}
\newcommand{\colbbonebinomktwo}{light blue}

\newcommand{\blind}{0}

\ifTAS
\fi

\begin{document}

\ifTAS
\else
\maketitle
\fi

\def\spacingset#1{\renewcommand{\baselinestretch}%
{#1}\small\normalsize} \spacingset{1}

\ifTAS
\if0\blind
{
  \title{\bf Flexible Bayesian Multiple Comparison Adjustment Using Dirichlet Process and Beta-Binomial Model Priors}
  \author{Don van den Bergh\\
    Department of Psychological Methods, University of Amsterdam\\
    and \\
    Fabian Dablander \\
    Institute for Biodiversity and Ecosystem Dynamics, University of Amsterdam\\
    Institute for Advanced Study, University of Amsterdam}
  \maketitle
} \fi

\if1\blind
{
  \bigskip
  \bigskip
  \bigskip
  \begin{center}
    {\LARGE\bf Flexible Bayesian Multiple Comparison Adjustment Using Dirichlet Process and Beta-Binomial Model Priors}
\end{center}
  \medskip
} \fi
\fi

\begin{abstract} 
\noindent Researchers frequently wish to assess the equality or inequality of groups, but this poses the challenge of adequately adjusting for multiple comparisons. Statistically, all possible configurations of equality and inequality constraints can be uniquely represented as partitions of groups, where any number of groups are equal if they are in the same \reviseAStwo{subset of the} partition. In a Bayesian framework, one can adjust for multiple comparisons by constructing a suitable prior distribution over all possible partitions. Inspired by work on variable selection in regression, we propose a class of flexible beta-binomial priors for multiple comparison adjustment. We compare this prior setup to the Dirichlet process prior suggested by \textcite{gopalan1998bayesian} and multiple comparison adjustment methods that do not specify a prior over partitions directly. 
Our approach not only allows researchers to assess pairwise \reviseAS{equality constraints} but simultaneously all possible equalities among all groups. Since the space of possible partitions grows rapidly --- for ten groups, there are already 115,975 possible partitions --- we use a stochastic search algorithm to efficiently explore the space. Our method is implemented in the Julia package \textit{EqualitySampler}, and we illustrate it on examples related to the comparison of means, \reviseAStwo{standard deviations}, and proportions.
\end{abstract}

\ifTAS
\noindent%
{\it Keywords:} Equality Constraints, Partitions, Stochastic Search, Markov Chain Monte Carlo

\newpage
\spacingset{1.45} 

\vfill
\else
\fi

\section{Introduction}
Assessing the equality or inequality of groups \reviseAStwo{(e.g., in terms of means, proportions, or standard deviations)} is a key problem in science and applied settings. If a confirmatory hypothesis is lacking, a standard approach is to first test whether all groups are equal and, if they are not, engage in multiple post-hoc comparisons. A large swathe of multiple comparisons techniques to guard against inflated false-positive errors exist in classical statistics \parencite[e.g.,][]{midway2020comparing}, dating back to the work of John Tukey and others \parencite[e.g.,][]{rao2009multiple, benjamini2002john}. From a Bayesian perspective, the problem of multiple comparisons can be addressed by changing the model prior \parencite[e.g.,][]{jeffreys1961theory, westfall1997bayesian, berry1999bayesian, debayesian2019}, an approach that has found prominent application in variable selection for regression \parencite[e.g.,][]{scott2006exploration, scott2010bayes}. Statistically, all possible configurations of equality and inequality constraints can be uniquely represented as partitions of the groups, where two groups are equal if they are in the same \reviseAStwo{subset of the} partition. In a Bayesian framework, one can adjust for multiple comparisons by constructing a suitable prior distribution over all possible partitions. This allows the researcher to explore the set of all possible equality and inequality relations among the groups while penalizing for multiple comparisons.

\revise{While there is a large body of work focusing on multiple hypothesis testing, that is, testing whether a location parameter is zero \parencite[e.g.,][]{dahl2007multiple, kim2009spiked, denti2021two, guo2010multiplicity, chang2020frequentist, bogdan2008comparison}, there has been considerably less attention to multiple comparison adjustments for testing the equality constraints among groups.} The first to propose a prior over all partitions to adjust for \revise{multiple comparisons} were, to our knowledge, \textcite{gopalan1998bayesian}, who suggested the Dirichlet process prior. \revise{This can be understood as a form of clustering on the level of parameters. In contrast to clustering the data, where the Dirichlet process prior is known to be inconsistent \parencite{miller2013simple}, using the Dirichlet process prior in the context of multiple comparison yields consistent estimates \parencite{quintana2003bayesian}.} 

\reviseAStwo{A related but distinct strand of recent work in Bayesian nonparametrics focuses on nested clustering structures to model grouped data. These models build on and generalize the nested Dirichlet process \parencite{rodriguez2008nested} by introducing flexible priors over random probability measures that induce clustering both within and between groups \parencite[e.g.,][]{beraha2021semi, d2024finite, lijoi2023flexible}. In doing so, they learn the number of clusters at both levels directly from the data. While \textcite{gopalan1998bayesian}'s approach and our own share the goal of identifying which groups can be treated as equal, they take complementary routes: rather than modeling the full data-generating distribution for each group, we assume a particular parametric form for data within each group and specify a prior directly on the space of partitions. This yields a framework that is interpretable and computationally efficient, particularly when interest lies in comparing summary statistics (e.g., means, variances, proportions) across groups rather than clustering both observations and groups. Specifically, we propose a class of flexible beta-binomial priors for Bayesian multiple comparison adjustment, inspired by work on variable selection in regression \parencite{scott2006exploration, scott2010bayes} and explore its properties in relation to previous work on multiple comparisons.}

The paper is structured as follows. In Section~\ref{sec:setup}, we set up the problem and describe the urn scheme from which a number of priors can be derived. We characterize three such priors --- the Dirichlet process, the beta-binomial, and the uniform prior --- and outline our methodology in Section~\ref{sec:methodology}. In Section~\ref{sec:simulation-study}, we contrast the priors, illustrate our method on a simulated example, and present a simulation study assessing the multiplicity adjustment of each prior. We also assess the method proposed by \textcite{westfall1997bayesian} and an uncorrected testing procedure based only on pairwise Bayes factors. As the space of possible partitions grows quickly --- for ten groups, there are already 115,975 possible partitions --- we set up a stochastic search algorithm to efficiently explore the space. Our method is implemented in Julia and available in the \textit{EqualitySampler} package from \url{https://github.com/vandenman/EqualitySampler.jl}. In Section~\ref{sec:applications} \reviseAStwo{and Appendix~\refAppStdev{}}, we apply our method to examples related to the comparison of proportions, \reviseAS{means}, and standard deviations. We conclude in Section~\ref{sec:discussion}.

\section{Preliminary Remarks} \label{sec:setup}
In this section, we set up the hypothesis testing problem, discuss the relation between partitions and models, and describe an urn scheme that will unify the presentation of the priors in the following section. 

\subsection{Problem Setup}
Our goal is to adjust for multiple comparisons in a flexible manner. Multiple comparisons are not a problem if we wish to compare only two hypotheses, denoted as $\mathcal{H}_0$ and $\mathcal{H}_1$. The Bayes factor quantifies how strongly we should update our prior beliefs about $\mathcal{H}_0$ relative to $\mathcal{H}_1$ after observing the data \parencite{kass1995bayes, ly2016harold}. Let group $j$ consist of $n_j$ observations $\vec{y}_j = \{y_{j1}, \ldots, y_{jn_j}\}$ for $j \in \{1, \ldots, K\}$ and $i \in \{1, \ldots, n_j\}$, and let $\vec{y} = \{\vec{y}_1, \ldots ,\vec{y}_K\}$. The Bayes factor is given by:
\begin{equation}
    \underbrace{\frac{p(\mathcal{H}_0 \mid \vec{y})}{p(\mathcal{H}_1 \mid \vec{y})}}_{\text{Posterior odds}} = \underbrace{\frac{p(\vec{y} \mid \mathcal{H}_0)}{p(\vec{y} \mid \mathcal{H}_1)}}_{\text{Bayes factor}} \, \, \times \underbrace{\frac{p(\mathcal{H}_0)}{p(\mathcal{H}_1)}}_{\text{Prior odds}} \enspace ,
\end{equation}
which does not depend on the number of hypotheses a researcher wishes to test.

A principled way to account for multiplicity is by adjusting the prior probability of the hypotheses \parencite[e.g.,][]{jeffreys1961theory, westfall1997bayesian}. Suppose a researcher is interested in comparing $K$ groups \reviseAS{in terms of the parameter of interest, denoted} by $\vec{\theta} = (\theta_1, \ldots, \theta_K)$. \reviseAS{These parameters can be anything the researcher wants to compare the groups on, such as proportions (see Section~\ref{sec:app:proportions}), means (see Section~\ref{sec:app:means}), or standard deviations (see Appendix~\refAppStdev)}. She is not only interested in whether all parameters are equal ($\mathcal{H}_0$) or whether they are \reviseAS{all} unequal ($\mathcal{H}_1$), but also which pairs of parameters are equal or not. In the language of classical statistics, she is interested in post-hoc comparisons. We focus on a Bayesian solution to this problem in the current paper. More specifically, going beyond classical testing, we consider the problem of assessing all possible equalities and inequalities between the groups. In general terms, the inference problem is:
\begin{align*}
    \rho &\sim \pi_{\rho}(.) \\
    \vec{\theta} \mid \rho &\sim \pi_{\vec{\theta}}(.) \\
    f(\vec{y}; \vec{\theta}, \rho) &= \prod_{j=1}^K g(\vec{y}_{j}; \theta_j, \phi) \enspace ,
\end{align*}
where $\rho$ is a partition, $\phi$ is a nuisance parameter (in case it exists), and $f$ and $g$ are the likelihood functions. Using the posterior distribution of $\vec{\theta}$, we have that:
\begin{align*}
    p(\mathcal{H}_0 \mid \vec{y}) &= p(\theta_1 = \theta_2 = \ldots = \theta_K \mid \vec{y}) \\
    p(\mathcal{H}_1 \mid \vec{y}) &= p(\theta_1 \neq \theta_2 \neq \ldots \neq \theta_K \mid \vec{y}) \enspace .
\end{align*}
\reviseAS{Note that while $\mathcal{H}_0$ and $\mathcal{H}_1$ denote the classical null and alternative hypothesis, respectively, t}here \reviseAS{exist} many more possible hypotheses, depending on the combination of equalities and inequalities. We can represent those as partitions, as we detail in the next section. 

\subsection{Partitions}
\begin{figure}
    \centering
    \includegraphics[width = \textwidth, keepaspectratio]{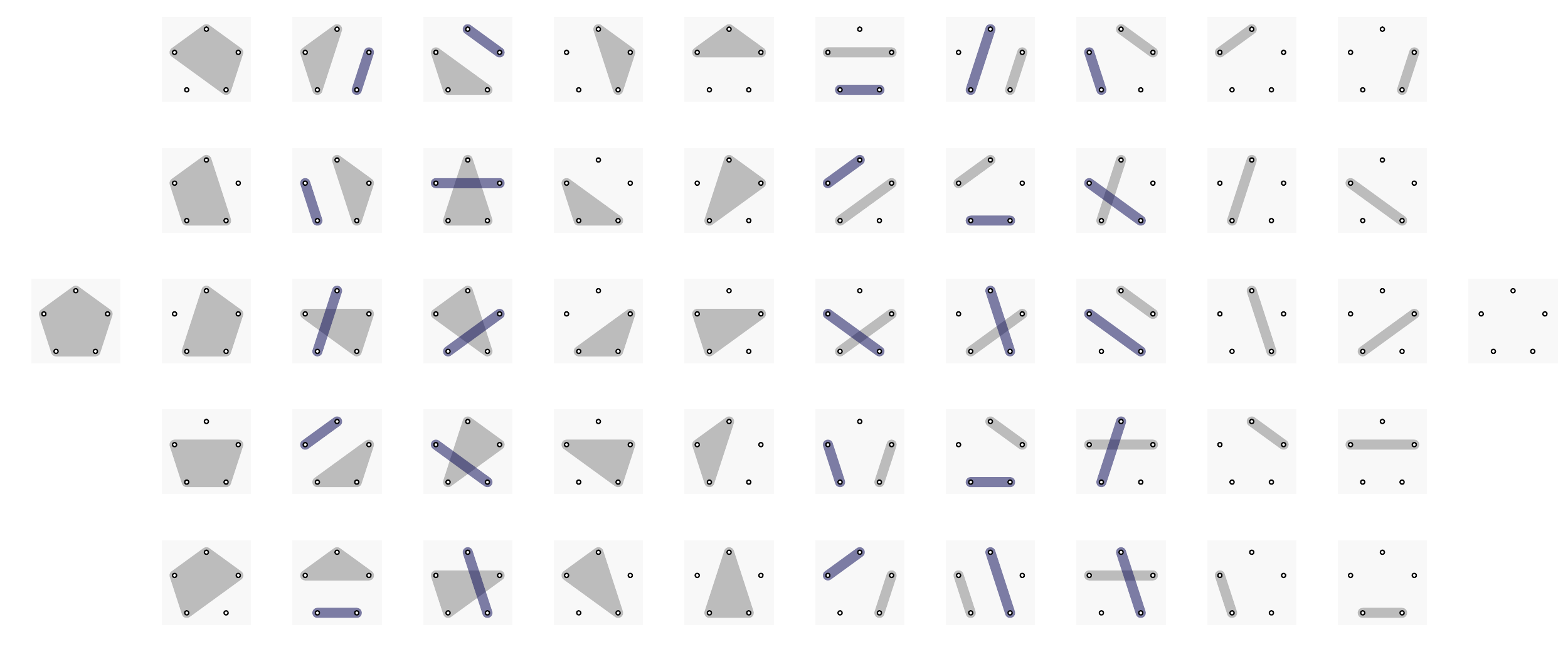}
    \caption{All 52 possible models given $K = 5$, represented as partitions. Circles represent individual parameters and shaded regions indicate which parameters are equal.}
    \label{fig:partitions}
\end{figure}

The space of possible equality constraints for some parameter vector $\vec{\theta} = (\theta_1, \ldots, \theta_K)$ of size $K$ is equivalent to the partitions of that vector. For example, for $K = 3$ the model that states $\theta_1 = \theta_2 \neq \theta_3$ is equivalent to the partition $\{\{\theta_1, \theta_2\}, \{\theta_3\}\}$. The space of possible models for $K = 5$ is shown in Figure~\ref{fig:partitions}. The correspondence between equality constraints and partitions is useful as partitions have been studied extensively in combinatorics. Given $K$ parameters, the number of partitions of size $j$ is given by the Stirling numbers of the second kind, denoted $\stirling{K}{j}$. The total number of partitions is given by the $K$\textsuperscript{th}-Bell number, which is defined as a sum over the Stirling numbers:
\begin{equation}
    \bellnum{K} = \sum_{j = 0}^K \stirling{K}{j} \enspace .
\end{equation}
The Bell numbers grow very quickly, with the number of partitions for a vector $\vec{\theta}$ of size 10 being $B_{10} = 115,975$.

The Stirling numbers and Bell numbers can be generalized to the $r$-Stirling \parencite{broder1984r} and $r$-Bell numbers \parencite{mezo2011r}, respectively. These generalizations help to construct conditional distributions, as \reviseAStwo{$r$ can be interpreted as the number of unique parameters}. The $r$-Stirling numbers $\rstirling{K}{j}{r}$ give the number of partitions of size $j$ given \reviseAStwo{$K$} groups such that the first $r$ parameters are all in distinct subsets \reviseAStwo{and are defined as:}
\begin{align}
    \reviseAStwo{\rstirling{K}{j}{r}} &\reviseAStwo{= \sum_{i=0}^K \binom{K}{i}\stirling{i}{j}r^{K-i}.}
\end{align}
\reviseAStwo{For example, suppose $K = 4$ and the first two parameters are known to be distinct, i.e., $r=2$. Then $\rstirling{4}{j}{2}$ for $j$ in $1,\,\dots,\,4$ counts the number of partitions with $1, \dots, K$ clusters, in this case $0$, $4$, $5$, and $1$.}
The $r$-Bell numbers give the total number of partitions given \reviseAStwo{$K+r$} parameters where the first $r$ parameters are in distinct subsets. \reviseAStwo{Thus, we may evaluate $\rbellnum{K-r}{r}$ to count all possible partitions of a parameter vector given that the first $r$ parameters are distinct, for example, $\rbellnum{4-2}{2} = 10$. The $r$-Bell number in this example equals the sum of the $r$-Stirling numbers in the previous example; this is also how the $r$-Bell numbers are often defined:}
\begin{align}
    \rbellnum{K}{r} &= \sum_{i=0}^K \rstirling{K+r}{i+r}{r} \enspace .
\end{align}
\reviseAStwo{These definitions hold for $K \geq j \geq r$, otherwise both numbers are defined to equal 0.}
Note that $\rstirling{K}{j}{1} = \stirling{K}{j}$ and that $\rbellnum{K}{0} = \bellnum{K}$. Both the $r$-Stirling and $r$-Bell numbers \reviseAStwo{can be computed} through recurrence relations, although explicit expressions exist which are easier to compute for large values; see \textcite{broder1984r} and \textcite{mezo2011r} for details.

\subsection{Urn Schemes}
We can \reviseAStwo{construct} the different partitions using an urn with $K$ different balls labeled 1 through $K$. For each parameter $\theta_j$, a ball $b_j$ is drawn \reviseAStwo{with replacement} from the urn with $b_j \in \{1, \ldots, K\}$. If two drawn balls are equal, $b_i = b_j$, then the two parameters are assigned to the same subset of the partition, that is, the two parameters $\theta_i$ and $\theta_j$ are equal if $b_i = b_j$. Note that different draws from an urn can represent the same partition. For example, the draws $(1, 1, 2)$ and $(3, 3, 1)$ both represent the partition $\{\{\theta_1, \theta_2\}, \{\theta_3\}\}$. The prior distributions introduced in the next sections assign probabilities to the unique partitions. Note that the prior probability of a particular draw can be obtained by dividing the probability of the corresponding partition by the total number of draws that correspond to that partition. The total number of draws that represent the same partition is given by $d!\binom{K}{d}$ where $d$ is the number of non-empty subsets of a particular draw.

Although the urn consists of $K$ different\reviseAStwo{ly labeled} balls, the event of interest is whether the next ball drawn equals one of the balls already drawn --- in other words, whether an equality or inequality is introduced. 
All prior distributions discussed below are related to \reviseAStwo{an urn scheme}.
Specifically, the joint prior distribution on $(\theta_1, \ldots, \theta_K)$ is \reviseAStwo{defined by the conditional distribution for $\theta_{j+1}$ given $\theta_1, \ldots, \theta_j$}:
\begin{revisionTwo}
\begin{align*}\label{eq:prediction-rule}
    \theta_1 &\sim \mathcal{K},\\
    \theta_{j+1} \mid \theta_1, \ldots, \theta_{j} &\sim \begin{cases}
    \mathcal{K} & \text{with probability } P_{\pi} \\
    \prior{\theta_1, \ldots, \theta_{j}} & \text{with probability }  1- P_{\pi} \enspace.
    \end{cases} \numberthis,
\end{align*}
\end{revisionTwo}
\reviseAStwo{which is also known as the \textit{prediction rule} \parencite[e.g.,][]{ishwaran2001gibbs}. 
Here, $\mathcal{K}$ denotes the base distribution from which novel realizations are drawn, and $\prior{\theta_1, \ldots, \theta_{j}}$ denotes a categorical distribution over previously seen values.
The specific prior distribution on partitions determines the probability $P_{\pi}$.
}
We characterize the priors we discuss in the next section in terms of (\ref{eq:prediction-rule})\reviseAStwo{,} in terms of the induced prior over partitions\reviseAStwo{,} and in terms of their penalty for multiplicity.

\section{Methodology} \label{sec:methodology}
Let $\vec{\theta^{\star}} = (\theta^{\star}_1, \ldots, \theta^{\star}_\len)$ denote the vector of unique population parameters out of $\vec{\theta} = (\theta_1, \ldots, \theta_K)$, $\vec{\theta}_{-j}$ the vector of parameters without parameter $\theta_j$, and the number of repeats of $\theta^{\star}_j$ as $n^{\star}_j$. 
\reviseAStwo{The number of unique parameters in a subvector $(\theta_1, \ldots, \theta_j)$ is given by $\lenj$ and the repeats by $n^{\star}_{\len_j}$.}
Let $\rho$ denote a partition and $|\rho|$ its size. For example, if $\rho = \{\{\theta_1, \theta_2\}, \{\theta_3\}\}$, then $|\rho| = 2$. Similarly, for this example $\vec{\theta^{\star}} = (\theta^{\star}_1, \theta^{\star}_2)$ and $n^{\star} = (2, 1)$. In the next sections, we first introduce and then contrast \reviseAStwo{the Dirichlet Process, the beta-binomial, and the uniform prior}. \reviseAStwo{Appendix~\refAppD{} discusses a natural generalization of the Dirichlet process prior, the Pitman–Yor process prior, but we found it offered no substantial improvement in multiplicity control.}

\subsection{Dirichlet Process Prior}
The Dirichlet process (DP) is a distribution over distributions \parencite{ferguson1973bayesian}. We say that $\mathcal{G} \sim \text{DP}(\alpha, \mathcal{K})$ is distributed according to a DP if its marginal distributions are Dirichlet distributed, where $\alpha$ is a concentration parameter and $\mathcal{K}$ is the base distribution, which will depend on the application; for details, see for example \textcite{teh2010dirichlet}. The DP can be understood as the infinite-dimensional generalization of the Dirichlet distribution, which makes it popular for mixture modeling \parencite[e.g.,][]{rasmussen1999infinite}. Our modeling approach is similar to mixture modeling, except that we do not cluster data but parameters --- a cluster corresponds to a partition. The prediction rule of the DP is given by \parencite[e.g.,][]{ishwaran2001gibbs, blackwell1973ferguson}:
\begin{align*} \label{eq:prediction-rule-DP}
    \reviseAStwo{\theta_1} &\reviseAStwo{\sim \mathcal{K},}\\
    \theta_{j + 1} \mid \theta_1, \ldots, \theta_j &\sim \begin{cases}
    \mathcal{K} & \text{with probability } \frac{\alpha}{\alpha + j} \\
    \text{Categorical}\left(\theta_1^{\star}, \ldots, \theta_\lenj^{\star} \mid n^{\star}_1, \ldots, n^{\star}_\lenj\right) & \text{with probability } 1 - \frac{\alpha}{\alpha + j} \enspace ,
    \end{cases} \numberthis
\end{align*}
where $\alpha$ is the concentration parameter and the base distribution of the DP depends on the application (see Section~\ref{sec:applications}). In other words, we draw a new value for $\theta_j$ from $\mathcal{K}$ with probability \reviseAStwo{$\nicefrac{\alpha}{\alpha + j}$}, or else set it to a previously observed value. \reviseAS{Larger values of $\alpha$ yield more \reviseAStwo{clusters,} and smaller values yield fewer \reviseAStwo{clusters.}} The particular value $\theta^{\star}_j$ the parameter $\theta_j$ is set to is proportional to the number of times $\theta^{\star}_j$ was observed previously, given by $n^{\star}_j$, resulting in the well-known ``rich get richer'' property \parencite[e.g.,][]{teh2010dirichlet}.

The Dirichlet process implies \reviseAS{the following} prior distribution over partitions \reviseAS{$\rho$:}
\begin{equation}
    \pi(\rho \mid \alpha) = \frac{\alpha^{|\rho|}\Gamma(\alpha)}{\Gamma(\reviseAStwo{K} + \alpha)} \prod_{c \in \rho} \Gamma(|c|) \enspace ,
\end{equation}
where $c$ is an element of $\rho$, and $|c|$ is its size. While the Dirichlet process features the infinite-dimensional object $\mathcal{K}$, the prior over partitions \reviseAS{is finite-dimensional}. \reviseAS{In other words, while the Dirichlet process itself is often introduced as a ``nonparametric'' prior — due to the unbounded number of potential mixture components — it ultimately induces a probability measure over partitions of, in our case, parameters that we wish to compare and that can be represented in a parametric form. This parametric form is known in the literature as an example of a \textit{product partition model}} \parencite[e.g.,][]{quintana2006predictive, quintana2003bayesian}, and makes it usable for our purposes, where we have a fixed number of parameters.

\begin{figure}[!ht]
    \centering
    \includegraphics[width = \textwidth]{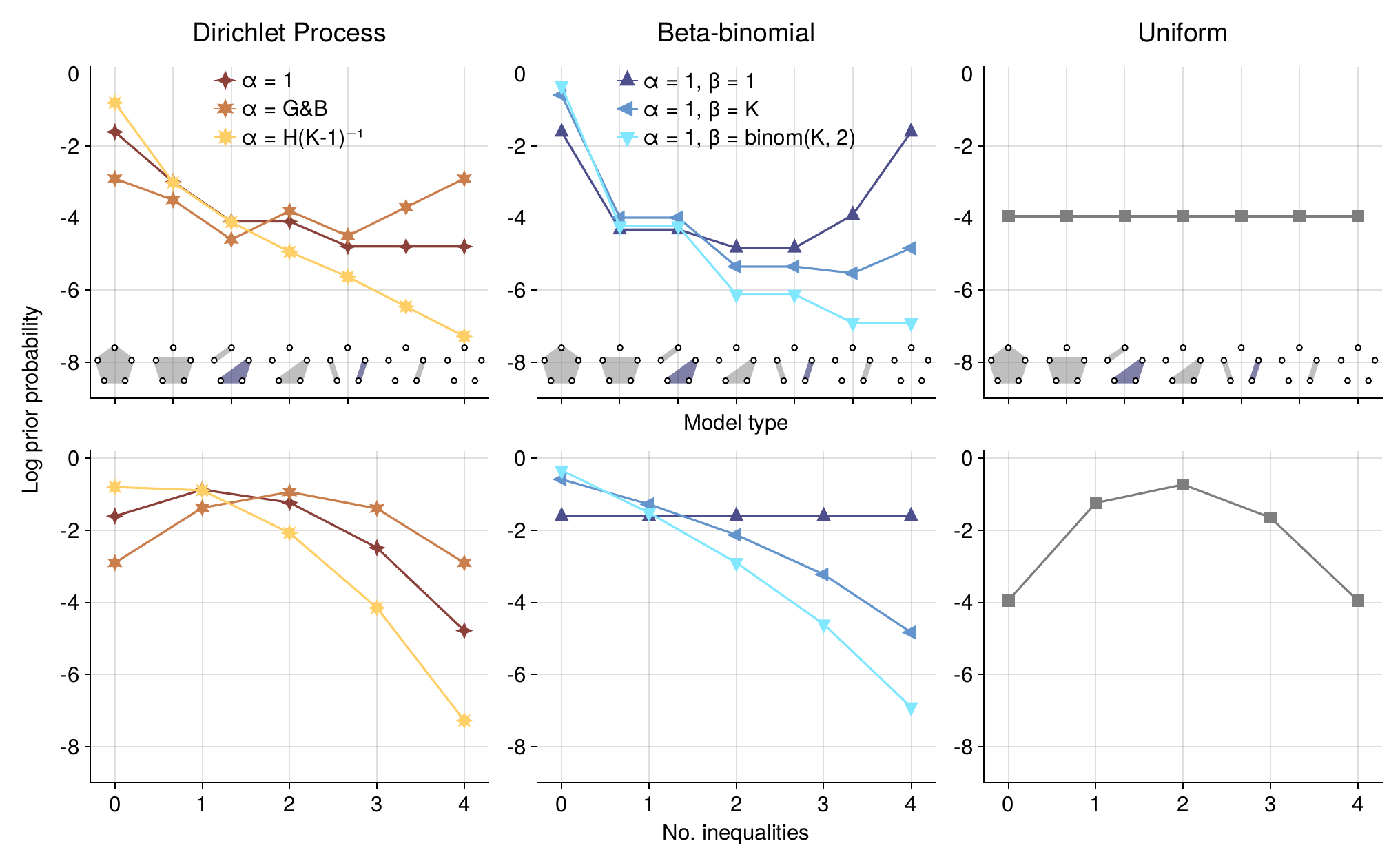}
    \caption{Top: Dirichlet process (left), beta-binomial (middle), and uniform prior (right) across distinct model types for $K = 5$ groups and different prior parameters. Bottom: Same but for the number of inequalities across models.
    }
    \label{fig:prior-comparison}
\end{figure}

The leftmost column in Figure~\ref{fig:prior-comparison} shows the DP prior over partitions (top) and number of inequalities (bottom) for different values of $\alpha$. \reviseAStwo{One requirement for a prior to penalize \textit{model complexity} is to be decreasing in the number of inequalities, which is distinct from penalizing \textit{multiplicity} (see Section \ref{sec:prior_comparison2})}. This is the case for \reviseAS{$\alpha = \nicefrac{1}{H_{K-1}}$ (yellow octagrams)} as shown in the top and bottom panels, and \revise{for any value $\alpha \leq \nicefrac{1}{H_{K-1}}$, where $H_n$ is the n\textsuperscript{th} harmonic number (see Appendix~\refAppDecrDpp)}. The value suggested by \textcite{gopalan1998bayesian} creates a symmetric prior over the partitions \reviseAS{(terracotta hexagrams)}, implying that the model with no inequalities is \textit{a priori} as likely as the model with all inequalities  (in the $K = 5$ case, this yields $\alpha = 2.213$). The prior with $\alpha = 1$ \reviseAS{(violet stars)} results in a nonincreasing prior over the number of partitions, but in an increasing prior over the number of inequalities: the model with one inequality is more likely than the model with no inequalities.

As $\alpha \rightarrow 0$, the prior of the model with all $K - 1$ equalities $\mathcal{M}_0$ (i.e., the null model) converges to one, while as $\alpha \rightarrow \infty$, the prior of the model with $K - 1$ inequalities $\mathcal{M}_{B_K}$ (i.e., the full model) converges to one. For prior elicitation, \textcite{gopalan1998bayesian} note that $\alpha$ is determined by specifying two of either $P(\mathcal{M}_0)$, $P(\mathcal{M}_{B_K})$, or their ratio, since $P(\mathcal{M}_0) = \nicefrac{\alpha (K - 1)!}{\prod_{j = 1}^K (\alpha + j - 1)}$ and $P(\mathcal{M}_{B_K}) = \nicefrac{\alpha^K}{\prod_{j = 1}^K (\alpha + j - 1)}$; see also Table~\ref{tab:overview}.

\begin{table}[!ht]
\begin{center}
\begingroup
\renewcommand{\arraystretch}{1.75}
\resizebox{\textwidth}{!}{%
\begin{tabular}{l ccc ccc ccc}
\toprule
 & \multicolumn{1}{c}{Dirichlet process prior} & \multicolumn{1}{c}{Beta-binomial prior} & \multicolumn{1}{c}{Uniform prior}\\
\midrule
Parameters  & $\alpha$ & $(\alpha = 1, \beta)$ & \xmark \\[1.4mm]
Prior over partitions  & $\frac{\alpha^{|\rho|}\Gamma(\alpha)}{\Gamma(n + \alpha)} \prod_{c \in \rho} \Gamma(|c|)$ & $\binom{K - 1}{|\rho| - 1}\frac{\FBeta{|\rho| - 1 + \alpha}{K - |\rho| + \beta}}{\FBeta{\alpha}{\beta}\stirling{K}{|\rho|}}$ & $\left(B_K\right)^{-1}$ \\[1.4mm]
Prior monotonically decreasing  & $\alpha \leq \nicefrac{1}{H_{K-1}}$ & $\beta \geq K$, $\beta \geq \binom{K}{2}$ & \xmark \\[1.4mm]
Prior probability of null model  & $\nicefrac{\alpha (K - 1)!}{\prod_{j = 1}^K (\alpha + j - 1)}$ & $\nicefrac{\FBeta{\alpha}{K - 1 + \beta}}{\FBeta{\alpha}{\beta}}$ & $\left(B_K\right)^{-1}$ \\[1.4mm]
Prior probability of full model & $\nicefrac{\alpha^K}{\prod_{j = 1}^K (\alpha + j - 1)}$ & $\nicefrac{\FBeta{K - 1 + \alpha}{\beta}}{\FBeta{\alpha}{\beta}}$ & $\left(B_K\right)^{-1}$ \\[1.4mm]
Prior odds of null / full model & $\nicefrac{\alpha (K - 1)!}{\alpha^K}$ & $\nicefrac{\FBeta{\alpha}{K - 1 + \beta}}{\FBeta{K - 1 + \alpha}{\beta}}$ & $1$ \\[1.4mm]
\reviseAS{%
Prior probability of ties $P(\theta_i = \theta_j)$
}& 
\reviseAStwo{%
$\nicefrac{1}{\alpha + 1}$
}& 
\reviseAS{%
$\sum_{i=1}^K BB(i|K, \alpha,\beta)\frac{\stirling{K-1}{i}}{\stirling{K}{i}}$
}
& 
\reviseAS{%
$\nicefrac{ \bellnum{K-1} }{ \bellnum{K} }$
}\\[1.4mm]
\bottomrule
\end{tabular}}
\endgroup
\end{center}
\caption{Characterizations of the different priors studied in this paper. Note: $\beta \geq K$ implies a prior decreasing in terms of the number of inequalities, but not in terms of the partitions. $\beta \geq \binom{K}{2}$ implies both.}
\label{tab:overview}
\end{table}

\subsection{Beta-binomial Prior}
The beta-binomial model prior is a popular choice for stochastic search variable selection in linear regression \parencite[][]{george1993variable} and Bayesian model averaging \parencite[e.g.,][]{hinne2020conceptual, hoeting1999bayesian}. It states that the prior probability of including $j$ predictors out of a total of $K$ predictors is given by:
\begin{equation}
    \BetaBinom{j}{K}{\alpha}{\beta} = \binom{K}{j} \frac{\FBeta{j + \alpha}{K - j + \beta}}{\FBeta{\alpha}{\beta}} \enspace ,
\end{equation}
where $\alpha$ and $\beta$ are hyperparameters \reviseAStwo{and $\FBeta{\alpha}{\beta}$ denotes the Beta function}. The prior probability of a particular regression model is obtained by dividing by the number of ways $j$ out of $K$ predictors can be included: $\BetaBinom{j}{K}{\alpha}{\beta} / \binom{K}{j}$. The beta-binomial distribution introduces a penalty for including additional predictors and in that way introduces a correction for multiplicity \parencite{scott2006exploration, scott2010bayes}. \reviseAS{For example, \textcite{scott2010bayes} show that for $\alpha = \beta = 1$, the prior probability of a model that includes no predictors is 30 times higher than the prior probability of a model that includes one predictor when testing a total of 30 variables. The penalty is less steep for other comparisons: a model that includes nine predictors is favored only by a factor of two compared to a model that includes ten predictors \parencite[for details, see][]{scott2010bayes}.} \revise{This beta-binomial prior over partitions can been seen as a special case of the hierarchical uniform prior where the distribution on the size of the partitions is a beta-binomial \parencite[]{casella2014cluster}.}

For the multiple comparison problem discussed in this paper, we consider the number of inequality constraints and use the beta-binomial prior to introduce a penalty for each additional inequality among the groups considered. For $K$ groups, there can be a maximum of $K - 1$ inequalities, resulting in a $\BetaBinom{i}{K-1}{\alpha}{\beta}$ prior distribution over the number of included inequalities $i$ out of $K$ groups. To see how this translates to a prior over the partitions $\rho$, note that there is a one-to-many correspondence between the number of inequalities $i$ out of $K$ groups and the resulting partitions $\rho$. For example, having $i = 1$ inequalities with $K = 3$ groups is consistent with the partitions $\{\{\theta_1, \theta_2\}, \{\theta_3\}\}$, $\{\{\theta_1, \theta_3\}, \{\theta_1\}\}$, and $\{\{\theta_2, \theta_3\}, \{\theta_1\}\}$, all of which are of size $|\rho| = i + 1$. The number of partitions of size $|\rho|$ is given, as discussed above, by the Stirling number $\stirling{K}{|\rho|}$. For the assignment of the prior probability, it is only the size of the partition (the number of inequalities) that counts. With these observations in hand, we arrive at the following (adjusted) beta-binomial prior distribution over partitions $\rho$:
\begin{equation}
    \pi(\rho \mid K, \alpha, \beta) = \binom{K - 1}{|\rho| - 1} 
    \frac{\FBeta{|\rho| - 1 + \alpha}{K - |\rho| + \beta}}
    {\FBeta{\alpha}{\beta}\stirling{K}{|\rho|}} \enspace .
\end{equation}
The prediction rule of the beta-binomial prior is given by:
\begin{align*} \label{eq:prediction-rule-BB}
    \reviseAStwo{\theta_1} &\reviseAStwo{\sim \mathcal{K},}\\
    \theta_{j + 1} \mid \theta_1, \ldots, \theta_j &\sim \begin{cases}
    \mathcal{K} & \text{with probability } P_{\pi} \\
    \text{Categorical}\left(\theta_1^{\star}, \ldots, \theta_\lenj^{\star} \mid 1, \ldots, 1\right) & \text{with probability } 1 - P_{\pi} \enspace .
    \end{cases} \enspace , \numberthis
\end{align*}
where
\begin{align}
    P_{\pi} &= 
    \frac{
        \sum_{\substack{\partition \in \Rho\\\theta_j \notin \vec{\theta}_{-j} \subseteq \rho}}
        \BetaBinom{\rho}{K}{\alpha}{\beta}
    }{
        \sum_{\substack{\partition \in \Rho\\\theta_j \notin \vec{\theta}_{-j} \subseteq \rho}}
        \BetaBinom{\rho}{K}{\alpha}{\beta} +
        \sum_{\substack{\partition \in \Rho\\\theta_j \in \vec{\theta}_{-j} \subseteq \rho}}
        \BetaBinom{\rho}{K}{\alpha}{\beta}\label{eq:11}
    } \enspace ,
\intertext{and where $\Rho$ denotes the set of all possible partitions. In essence, Equation \eqref{eq:11} takes the probability of all possible partitions where $\theta_j$ is distinct from $\vec{\theta}_{-j}$, conditional on $\vec{\theta}_{-j}$ being a subset of the considered partition. The sum over all possible partitions can be simplified using the $r$-Stirling numbers:}
    P_{\pi} &=
    \frac{
        \sum_{i=1}^K \BetaBinom{i}{K}{\alpha}{\beta} \rstirling{K-j+\lenj+1}{i}{\lenj + 1}
    }{
        \lenj \sum_{j=1}^K \BetaBinom{i}{K}{\alpha}{\beta} \rstirling{K-j+\lenj{}}{  i}{\lenj{} } +
              \sum_{i=1}^K \BetaBinom{i}{K}{\alpha}{\beta} \rstirling{K-j+\lenj{}+1}{i}{\lenj{} + 1}
    } \enspace ,
\end{align}
where $\lenj$ is \reviseAStwo{the} number of unique parameters in \reviseAStwo{$\theta_1, \ldots, \theta_j$, that is, the number of clusters sampled so far.
A proof that the prediction rule leads to the desired joint distribution is given in Appendix~\refPredictionRuleProof.
}

The beta-binomial prior on the partitions and the induced prior on the number of inequalities are shown for different parameterizations in the middle column in Figure~\ref{fig:prior-comparison}. For $\alpha = \beta = 1$, the beta-binomial distribution over the partitions has a characteristic U-shape \reviseAS{(dark blue upward triangles)}. This prior specification in turn implies a uniform prior on the number of inequalities. We follow \textcite{wilson2010bayesian} who, in the context of regression, suggested to set $\alpha = 1$ as a default so that the distribution over model size (here the number of inequalities) is nonincreasing, and to scale $\beta = \lambda K$ with the number of groups to force the prior to be monotonically decreasing, with a default of $\lambda = 1$ \parencite{wilson2010bayesian}. This is illustrated as the \reviseAS{blue line} \reviseAS{(leftward triangles)} in Figure~\ref{fig:prior-comparison} using $\beta = 5$. In the multiple comparison case, we additionally investigate $\beta = \binom{K}{2}$, which implies that the prior on the number of inequalities of individual models is nonincreasing, see Appendix~\refAppDecrBb. The \reviseAS{light blue line} \reviseAS{(downward triangles)} in Figure~\ref{fig:prior-comparison} shows a decreasing prior for $\beta = \binom{5}{2} = 10$. This prior assigns the least mass to models with an increasing number of inequalities compared to all other beta-binomial priors.

Figure~\ref{fig:prior-comparison} shows that the DP prior makes a distinction that the beta-binomial is, by design, not making: while the beta-binomial prior assigns the same prior mass to partitions with the same number of equalities, the DP prior assigns more mass to the partition with the larger cluster. For example, the beta-binomial does not distinguish between $\{\{\theta_1, \theta_2, \theta_3\}, \{\theta_4\}, \{\theta_5\}\}$ and $\{\{\theta_1, \theta_2\}, \{\theta_3, \theta_4\}, \{\theta_5\}\}$, while the DP assigns more mass to the former (see Figure~\ref{fig:prior-comparison}). We return to this distinction in \reviseAStwo{Section \ref{sec:prior_comparison2} and} the discussion.

Lastly, note that for the beta-binomial prior we have that $P(\mathcal{M}_0) = \nicefrac{\FBeta{\alpha}{K - 1 + \beta}}{\FBeta{\alpha}{\beta}}$ and $P(\mathcal{M}_{B_K}) = \nicefrac{\FBeta{K - 1 + \alpha}{\beta}}{\FBeta{\alpha}{\beta}}$. Fixing $\alpha = 1$, we have that as $\beta \rightarrow \infty$, the prior of the model with all $K - 1$ equalities $\mathcal{M}_0$ converges to one, while as $\beta \rightarrow 0$, the prior of the model with $K - 1$ inequalities $\mathcal{M}_{B_K}$ converges to one; see also Table~\ref{tab:overview}. As with the Dirichlet process prior discussed above, one can use these relations in prior elicitation.

\subsection{Uniform Prior} \label{app:uniform}
For completeness, we give a prior that is uniform over the space of partitions. The probability mass function is straightforward, as all valid configurations of size $K$ have probability $\nicefrac{1}{\bellnum{K}}$. The prediction rule of the uniform prior is given by:
\begin{align*}
    \reviseAStwo{\theta_1} &\reviseAStwo{\sim \mathcal{K},}\\
    \theta_{j + 1} \mid \theta_1, \ldots, \theta_j
    &\sim \begin{cases}
    \mathcal{K} & \text{with probability } P_{\pi_{U}} \\
    \text{Categorical}\left(\theta_1^{\star}, \ldots, \theta_\lenj^{\star} \mid 1, \ldots, 1\right) & \reviseAS{\text{with probability } 1 - P_{\pi_{U}}} \enspace ,
    \end{cases} \numberthis
\end{align*}
where
\begin{align}
    P_{\pi_{U}} 
    &=
    \frac{\rbellnum{K-j}{\lenj + 1}}{\rbellnum{K-j}{\lenj + 1} + \lenj\rbellnum{K-j}{\lenj}} \enspace .
\end{align}
Here, $\rbellnum{K-j+1}{\lenj + 1}$ counts the number of models where $\theta_{j+1} \notin \left(\theta_1^{\star}, \ldots, \theta_\lenj^{\star}\right)$ conditional on $\theta_1, \ldots, \theta_j$ being assigned to $\lenj$ distinct subsets.
Complementarily, $\rbellnum{K-j+1}{\lenj}$ counts the number of models where $\theta_{j+1} \in \left(\theta_1^{\star}, \ldots, \theta_\lenj^{\star}\right)$ conditional on $\theta_1, \ldots, \theta_j$ being assigned to $\lenj$ distinct subsets, which is multiplied by $\lenj$ as there are $\lenj$ subsets that $\theta_{j+1}$ could be assigned to. Under this uniform prior, all partitions $\rho$ are equally likely, as can be seen in the top right panel in Figure~\ref{fig:prior-comparison}. Note that this uniform prior induces a non-uniform prior on the number of inequalities, as shown in the bottom right panel.

\subsection{Contrasting the Three Priors}\label{sec:prior_comparison2}
\reviseAStwo{Before contrasting the three priors, we clarify the distinction between penalizing model complexity and adjusting for multiple comparisons. Complex models have more parameters than simple models, and Bayesian methods penalize this through marginal likelihoods when integrating over the prior — a mechanism sometimes referred to as the Bayesian Ockham’s razor effect \parencite[cf.][]{scott2010bayes}. As \textcite{scott2010bayes} emphasize, multiplicity adjustment in the Bayesian context arises not from the integration over the parameter space, but from the specification of prior probabilities over the model space: a method adjusts for multiplicity when it allocates less prior probability to individual alternative models as the number of possible alternatives increases. In our context, this means that priors over partitions that increasingly concentrate mass on fewer inequality constraints — as the number of groups increases — help control the risk of overinterpreting differences due to random noise, that is, adjusts for multiple comparisons.}

\begin{figure}[!ht]
    \centering
    \includegraphics[width = \textwidth]{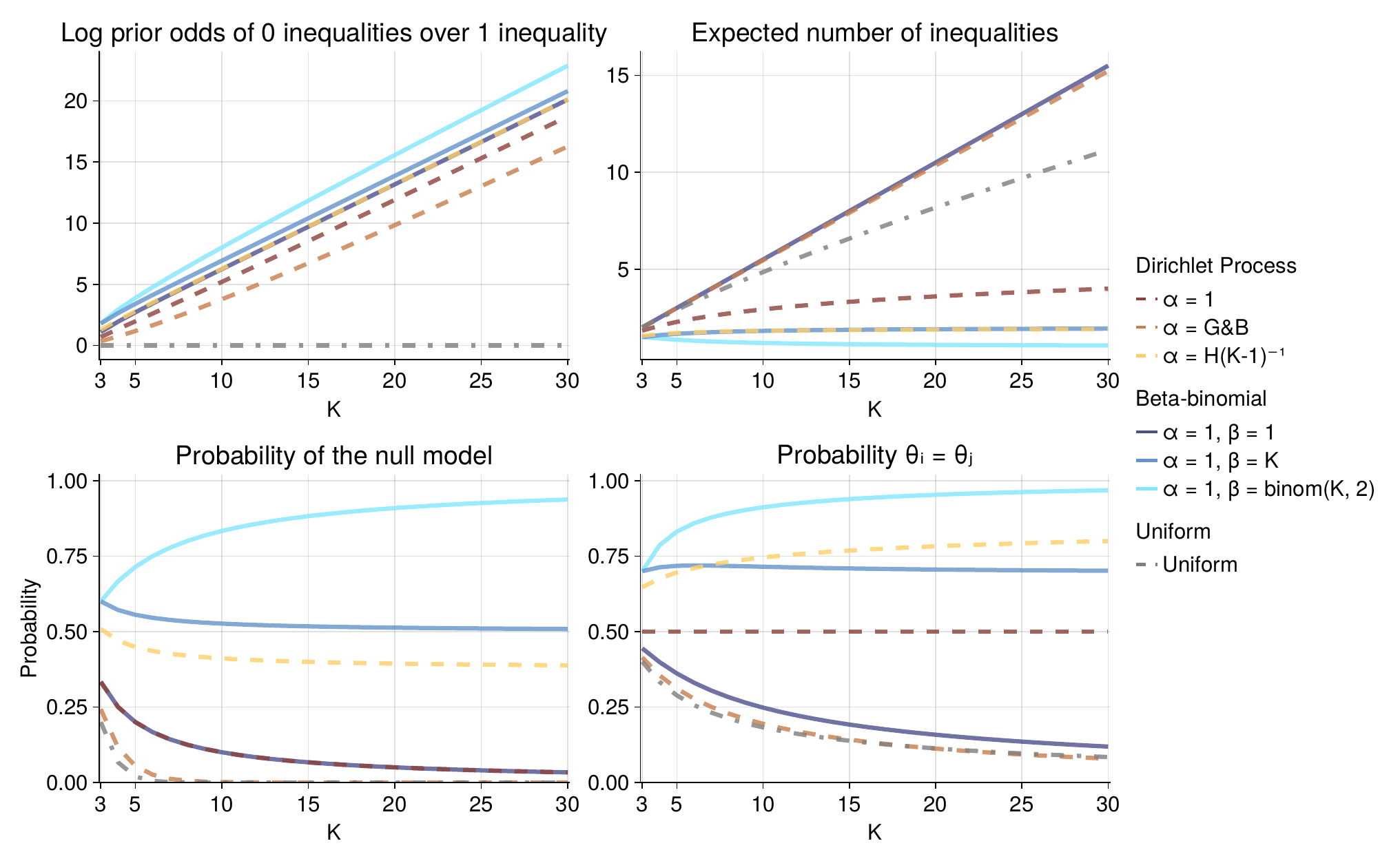}
    \caption{\reviseAStwo{The average log prior odds of a partition with 0 inequalities vs. 1 inequality (top left), expected number of clusters (top right)}, probability of the null model (\reviseAStwo{bottom} left), \reviseAStwo{and} probability of pairwise equality (\reviseAStwo{bottom right}), for various Dirichlet process and beta-binomial priors and the uniform prior for increasing numbers of groups $K$.}
    \label{fig:prior-comparison2}
\end{figure}

\reviseAStwo{We can see the extent to which different the priors penalize multiplicity in the top left panel of Figure~\ref{fig:prior-comparison2}, which shows the average log prior odds for a model with no inequalities versus models with one equality as the number of groups $K$ increases. We find that the $\text{BB}(1, \beta = \binom{K}{2})$ prior (\colbbonebinomktwo) shows the strongest multiplicity penalty, followed by the $\text{BB}(1, \beta = K)$ prior (\colbbonek), the $\text{BB}(1, \beta = 1)$ prior (dark blue), and the $\text{DP}(\alpha = \nicefrac{1}{H_{K-1}})$ prior (yellow), the latter two of which show the same penalty as $K$ increases. The $\text{DP}(\alpha = 1)$ prior (terracotta) and especially the Dirichlet process with the specification due to \textcite{gopalan1998bayesian} (violet), \reviseAStwo{$\text{DP}(\alpha = \text{G\&B})$}, exhibit considerably weaker multiplicity penalties, while the uniform prior (\coluniform) exhibits no penalty at all.}

The other panels provide further intuition for the differences between the priors, showing the \reviseAStwo{expected number of inequalities (top right), the probability of the null model (bottom left), and the probability of $\theta_i = \theta_j$ (bottom right)} as the number of groups $K$ increases. The $\text{BB}(1, \beta = \binom{K}{2})$ prior and the $\text{DP}(\alpha = \nicefrac{1}{H_{K-1}})$ prior both assign monotonically decreasing probabilities to the number of inequalities. The prior probability of the null model for the former model grows and approaches one as $K$ increases, while it approaches a constant for the latter. The probability that $\theta_i = \theta_j$ grows for both these priors as $K$ increases, with the beta-binomial prior exhibiting an almost constant higher value. Both priors naturally expect a very small number of \reviseAStwo{inequalities}, which for the beta-binomial prior approaches one while it approaches a slightly higher value for the Dirichlet process. The $\text{BB}(1, \beta = K)$ prior expects virtually the same number of \reviseAStwo{inequalities} as the $\text{DP}(\alpha = \nicefrac{1}{H_{K-1}})$ prior, and they seem to converge to similar values for the probability of the null model and the probability that any two parameters are equal. The $\text{BB}(1, \beta = 1)$ prior exhibits the same pattern for the probability of the null model as the $\text{DP}(\alpha = 1)$ prior. However, the Dirichlet process assigns a constant probability to pairwise equality of $\nicefrac{1}{2}$, while the beta-binomial prior assigns a probability that approaches zero with increasing $K$. Both the uniform prior and \reviseAStwo{$\text{DP}(\alpha = \text{G\&B})$} show the most rapidly decreasing probabilities for the null model and pairwise equalities. The uniform prior expects a much larger number of \reviseAStwo{inequalities} as $K$ increases, however.

What can we learn from this comparison? \reviseAStwo{First, the strongly increasing expected number of inequalities as $K$ increases rules out the $\text{BB}(1, \beta = 1)$ and \reviseAStwo{$\text{DP}(\alpha = \text{G\&B})$} as suitable priors. Second, while all priors except the uniform penalize multiplicity (and which is therefore ruled out), the $\text{BB}(\beta = \binom{K}{2})$ prior does so very strongly. For example, for $K = 10$ the average prior odds in favor of no inequalities is about 2981, while it is only about 1022 for the $\text{BB}(\beta = K)$ prior. The $\text{BB}(\beta = \binom{K}{2})$ prior \reviseAStwo{similarly} favors the null model too strongly, rendering it unsuitable. The $\text{DP}(\alpha = 1)$ prior exhibits sensible behavior, elegantly fixing the probability that $\theta_i = \theta_j$, but might allocate too little prior probability to the null model and penalize multiplicity too weakly. Overall, our favored priors are the $\text{BB}(\beta = K)$ and $\text{DP}(\alpha = \nicefrac{1}{H_{K-1}})$ prior, which exhibit very similar behavior and which we deem the most suitable for the multiple comparison problem from the set of priors studied here.}

Lastly, the priors differ conceptually in that the Dirichlet process prior includes preferential attachment, that is, new values $\theta^{\star}$ are more likely to be assigned to larger clusters, while the beta-binomial prior assigns the new value to existing clusters uniformly. In practice, however, there may not be a big difference between the beta-binomial prior and the Dirichlet process prior if one suitably aligns their prior specifications. Our contribution in this work is thus not to clearly crown one prior as ``the winner'', but study their behavior with attention to their differences and similarities. Before we turn to an extensive simulation study to do exactly that in Section~\ref{sec:simulation}, we detail our stochastic search method.

\subsection{Stochastic Search Method} \label{sec:method-description}
When the number of groups is small and the computation of Bayes factors is swift, one can directly compute the Bayes factors for all hypotheses. Using the priors we outlined above, one can then obtain posterior distributions over hypotheses that incorporate the desired multiplicity adjustment. The number of \reviseAS{possible equality constraints} grows extremely quickly with the number of groups, however, and for larger number of groups one must rely on stochastic search methods. Moreover, while directly computing the Bayes factors results in posterior distributions over hypotheses, it does not yield posterior distributions over parameters. We therefore set up a stochastic search method that yields both, allowing researchers to incorporate uncertainty across hypotheses through model averaging \parencite[e.g.,][]{hinne2020conceptual, hoeting1999bayesian}. 

Our method is implemented in the programming language Julia \parencite{Julia2017Bezanson}. First, we implemented the prior distributions in Julia. 
\begin{revision}
For the ANOVA model and proportion example, we integrate out all parameters except for the partitions and directly sample from the partition space.
To explore the partition space, we use two sampling steps.
The first sampling step is a local move; it deterministically enumerates all parameters and for each parameter, it proposes to move that parameter to any existing subset of the current partition or to a new set.
For example, given $\{\{\theta_1\}, \{\theta_2, \theta_3\}, \{\theta_4\}\}$ \reviseAStwo{and when updating $\theta_1$}, we propose to move to $\{\{\theta_1\}, \{\theta_2, \theta_3\}, \{\theta_4\}\}$, $\{\{\theta_1, \theta_2, \theta_3\}, \{\theta_4\}\}$, $\{\{\theta_1, \theta_4\} \text{ or } \{\theta_2, \theta_3\}\}$.
This boils down to a draw from a categorical distribution across partitions, with probabilities proportional to the integrated marginal likelihood evaluated for each partition.

The local move works well when the number of groups is limited, but when the number is substantial, as in Example~\ref{sec:app:means}, it can be slow to converge \parencite[as also noted in][]{miller2018mixture}. 
Therefore, as a second sampling step, we make global moves using an informed split-merge proposal. 
The split-merge proposal works as follows.
With probability \nicefrac{1}{2}, we either merge or split a subset of a partition.
If a merge is proposed, we compute all pairwise distances among the current subsets using an informed distance metric, for example the difference between the means of the subsets.
Next, we sample a pair of subsets to merge, where the probability of each pair is proportional to the inverse of the distance metric.
The split move similarly uses a distance metric to pick a subset to split, for example, the variance of each subset. 
Next, we use the median of the distance metric of the individual groups (e.g., the subgroup means) to divide the subset into two sets.
For each set, we compute the overall mean and variance and construct a \reviseAStwo{Gaussian} distribution corresponding to each set, say $\mathcal{N}_1$ and $\mathcal{N}_2$.
We compute the density of each parameter in the set under both $\mathcal{N}_1$ and $\mathcal{N}_2$, normalize these to probabilities, and then sample the assignment to each set.
This last part makes the split proposal probabilistic and ensures that any merge move is reversible.
\end{revision}

\section{Investigating Multiplicity Adjustment} \label{sec:simulation-study}
In this section, we investigate the differences between the above priors in more detail and compare them to the method proposed by \textcite{westfall1997bayesian} and an uncorrected approach using pairwise Bayes factors. In Section~\ref{sec:illustration}, we use a small simulation study to illustrate the implications of multiplicity adjustment. In Section~\ref{sec:simulation}, we present the results of a more extensive simulation study.

\subsection{Illustrating Multiplicity Adjustment} \label{sec:illustration}
Here we illustrate the different multiplicity penalties that the different priors impose using a small simulation study. We simulate data from a one-way ANOVA model and analyze it using the specification by \textcite{rouder2012default}. The ANOVA model extended with a prior over partitions is given by:
\begin{align*}
    Y_{ij}              &\sim \mathcal{N}\left(\mu + \sigma\theta_j, 1\right)\\
    \mu                 &\propto 1  \\
    \sigma^2            &\propto 1 / \sigma^2 \\
    g                   &\sim \mathcal{IG}\left(\nicefrac{1}{2}, \nicefrac{1}{2}\right) \\
    \vec{\theta}^u   &\sim \mathcal{N}_{\setsize{\partition}}\left(0, g\right)   \\
    \vec{\theta}^c   &\leftarrow \mathbf{Q}\vec{\theta}^u \\
    \theta_j            &\leftarrow \theta^c_l\text{ such that } j \in \rho_l\\
    \rho                &\sim \pi_{\rho}(.) \enspace , \numberthis
\end{align*}
\reviseAS{where $\mu \propto 1$ indicates an improper prior and $\sigma^2 \propto 1 / \sigma^2$ indicates Jeffreys's prior, both of which are routinely applied for testing purposes \parencite[e.g.,][]{ly2016harold}.} The data follow a Gaussian distribution with a grand mean $\mu$ and a group-specific offset $\theta_j$. The offsets sum to zero to avoid identification constraints. This is achieved by projecting $\vec{\theta}^u$ from a \reviseAStwo{$\setsize{\rho}-1$} dimensional space onto a \reviseAStwo{$\setsize{\rho}$} dimensional space using the matrix $\mathbf{Q}$, which consists of the first \reviseAStwo{$\setsize{\rho}-1$} columns of an eigendecomposition of a degenerate covariance matrix as defined in \textcite{rouder2012default}.\footnote{Note that this projection is not unique. It can also be achieved with, for example, a QR decomposition, as recommended by the \textcite{stanUserManual}.} 
\reviseAStwo{Next, $\vec{\theta}$ is obtained by matching the indices with the partition, e.g., for  $\rho = \{\{1,3\},\{2\}\}$ and $\vec{\theta}^c = \{-0.5, 0.5\}$ we obtain $\vec{\theta}=\{-0.5, 0.5, -0.5\}$.}
The unconstrained offsets $\vec{\theta}^u$ are assigned a $g$ prior where $g$ itself is assigned an inverse gamma prior with shape and scale equal to $\nicefrac{1}{2}$ \parencite{liang2008mixtures}. Note that the model reduces to the approach of \textcite{rouder2012default} whenever the partition indicates that all elements are distinct.


We simulated from the null model, which assumes that all the groups are equal, and from the full model, which assumes that all groups are unequal, drawing 100 observations per group and varying the number of groups $K \in [2, 3, \dots, 10]$, repeating each combination \reviseAS{200} times.
\reviseAS{In the full model, the means were of increasing size with successive differences, such that the differences between adjacent groups always equal the same number. To minimize the effect of increasing the number of groups, we fixed the average pairwise difference across different values of $K$ to $0.20$.} 
%
For the analysis, we considered \reviseAS{seven} priors: \reviseAStwo{the Dirichlet process priors $\text{DP}(\alpha = 1)$, $\text{DP}(\alpha = \text{G\&B})$, and $\text{DP}(\alpha = \nicefrac{1}{H_{K-1}})$; the beta-binomial priors $\text{BB}(\alpha = 1, \beta = 1)$, $\text{BB}(\alpha = 1, \beta = K)$, and $\text{BB}(\alpha = 1, \beta = \binom{K}{2})$}; and the uniform prior. We also included the prior adjustment method proposed by \textcite{westfall1997bayesian} and an uncorrected method using pairwise Bayes factors.

To assess how well the respective priors adjust for multiplicity, we calculated how frequently the posterior probability that any two groups differ is larger than 0.50, using the null model as data-generating model. Similarly, to assess how well the respective priors are capable of detecting true differences, we calculated how frequently the posterior probability that any two groups \textit{do not} differ is larger than 0.50, using the full model as data-generating model.

\begin{figure}
    \centering
    \includegraphics[width=\textwidth]{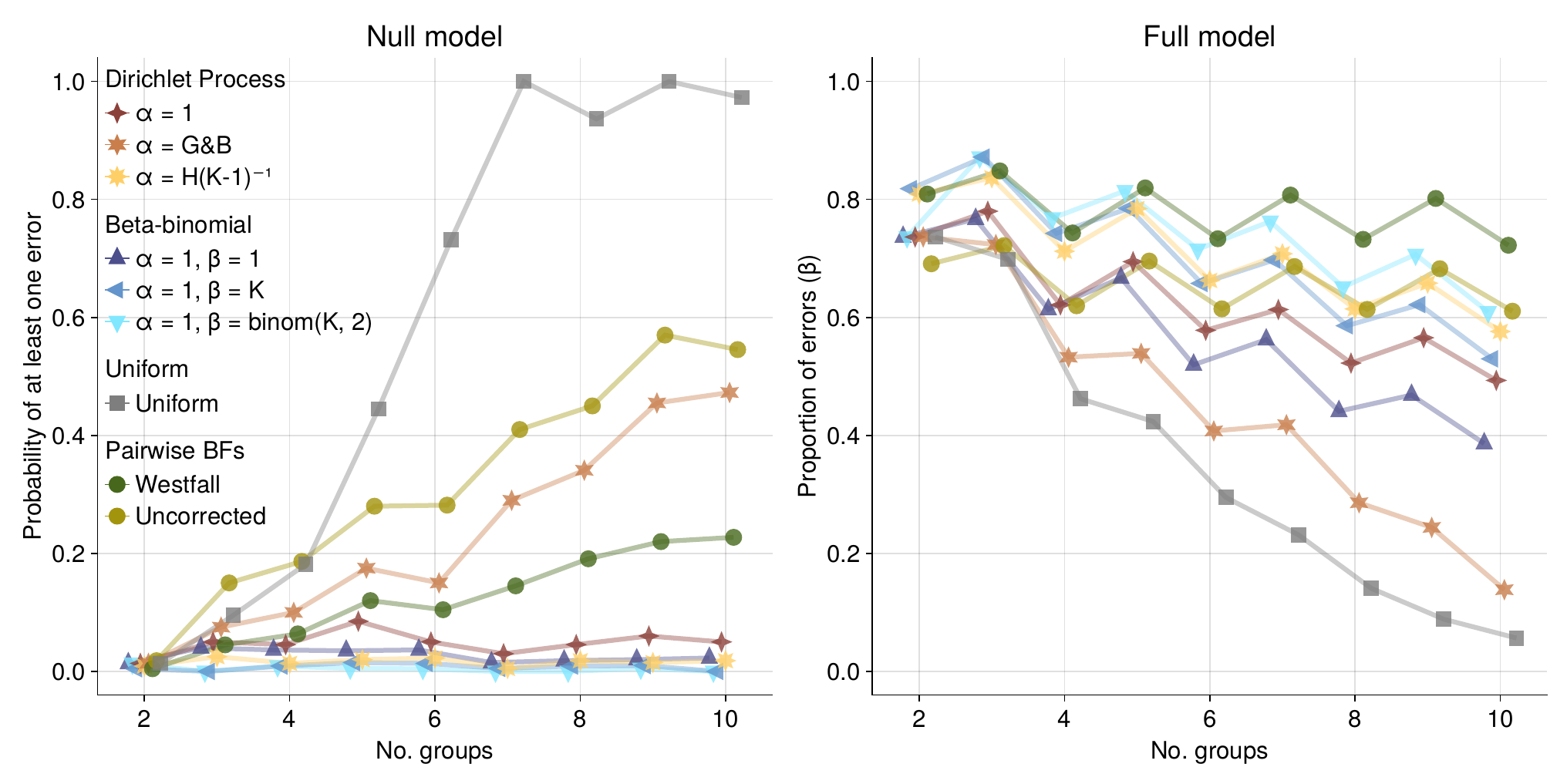}
    \caption{Left: Probability of making at least one false claim about a difference between two groups when there is none. Right: Proportion of falsely claiming no difference between two groups when there is one. 
    }
    \label{fig:small_simulation}
\end{figure}


The left panel in Figure~\ref{fig:small_simulation} shows that using a uniform prior (\coluniform \, squares) very quickly leads to false positives as the number of groups increases. This is not surprising: the uniform prior assigns each model the same prior mass, hence diminishing the plausibility assigned to $\mathcal{H}_0$ dramatically as $K$ increases, thus increasing the probability of an error.

\reviseAStwo{$\text{DP}(\alpha = \text{G\&B})$} \reviseAS{(terracotta hexagrams)} performs better than the uniform prior but still does not provide adequate error control. It performs roughly as poorly as the method which simply computes pairwise Bayesian $t$-tests \reviseAS{(light green circles)}. The correction proposed by \textcite{westfall1997bayesian} performs much better \reviseAS{(dark green circles)} but still leads to a relatively high probability of making at least one error as the number of groups increases. \reviseAStwo{$\text{DP}(\alpha = 1)$ \reviseAS{(violet stars)} performs better, with $\text{DP}(\alpha = \nicefrac{1}{H_{K-1}})$ (yellow octagrams)} and the set of beta-binomial priors providing good error control.

The right panel in Figure~\ref{fig:small_simulation} shows that the \reviseAS{uniform prior} leads to the lowest proportion of falsely claiming no difference between two groups, followed by \reviseAStwo{$\text{DP}(\alpha = \text{G\&B})$} and \reviseAStwo{$\text{BB}(\alpha = 1, \beta = 1)$}. The method proposed by \textcite{westfall1997bayesian} performs worst, followed by \reviseAStwo{$\text{BB}(\alpha = 1, \beta = \binom{K}{2})$} and \reviseAStwo{$\text{DP}(\alpha = \nicefrac{1}{H_{K-1}})$}. The performance of the uncorrected pairwise Bayes factor approach and \reviseAStwo{$\text{BB}(\alpha = 1, \beta = K)$} is somewhere in the middle. Note that all approaches perform better as the group size increases, but this is due to our simulation design: each additional group adds $n$ more observations, which makes falsely claiming no difference less likely with an increasing number of groups. Overall, we conclude that not adjusting for multiple comparisons — either by using a uniform prior or by using pairwise Bayes factors — naturally leads to the worst performance and that the method by \textcite{westfall1997bayesian} is overly conservative and does not provide adequate error control with an increasing number of groups. In the next section, we report on a more extensive simulation study to further disentangle the differences between the \reviseAS{priors}.

\subsection{Simulation Study} \label{sec:simulation}
In the previous section, we illustrated the importance of adjusting the prior model probabilities in reducing the familywise error rate when all groups are equal. Here we explore the multiplicity adjustment of the different methods in a more exhaustive simulation study. We used the same ANOVA model as in the previous section and varied the total number of groups $K \in \{5, 9\}$ and the sample size per group \reviseAS{$n \in \{50, 100, 200, 300, 400, 500\}$}. In addition, we varied the true number of equalities to be $\{0\%, 25\%, 50\%, 75\%, 100\%\}$. For $K = 5$, there are 4 possible \reviseAS{equality constraints} which resulted in models that have either 0, 1, 2, 3, or 4 equalities. For $K = 9$, there are 8 possible \reviseAS{equality constraints}, resulting in 0, 2, 4, 6, or 8 equalities in the true model. Given the number of equalities, we sampled a particular partition uniformly from all possible partitions with that amount of equalities and used this model to simulate data. Each unique combination was repeated \reviseAS{500} times and each generated data set was analyzed with the same prior specifications as above. We assessed the familywise error control as well as statistical power. The results for K = 5 and K = 9 were similar. Therefore, we focus on the $K = 5$ in the main text and discuss the $K = 9$ case in Appendix~\refAppExtraSim.

Note that the hierarchical approach has an additional source of $\alpha$ error in contrast to pairwise comparisons when there are more than 0 inequalities because it imposes transitivity. For example, imagine that the true model postulates that $\theta_1 = \theta_2 = \theta_3 \neq \theta_4$. However, the sample means are (by random sampling) $\bar{x}_1 = 0.1, \bar{x}_2 = 0.2, \bar{x}_3 = 0.3, \bar{x}_4 = 0.35$. The hierarchical approach would find that $\theta_3 = \theta_4$, but not that $\theta_1 = \theta_3$ since that also implies $\theta_1 = \theta_4$. Therefore, the model $\theta_1 = \theta_2 = \theta_3 \neq \theta_4$ and even the equality $\theta_1 = \theta_2$ are not retrieved. In contrast, the pairwise methods violate transitivity as they only look at two pairs at the time and will happily suggest that $\theta_1 = \theta_2$, $\theta_2 = \theta_3$, and $\theta_3 = \theta_4$ while simultaneously suggesting that $\theta_1 \neq \theta_4$.


\subsubsection{Familywise Error Rate} \label{sec:simulation-family}
\begin{figure}[!ht]
    \centering
    \includegraphics[width=1\textwidth]{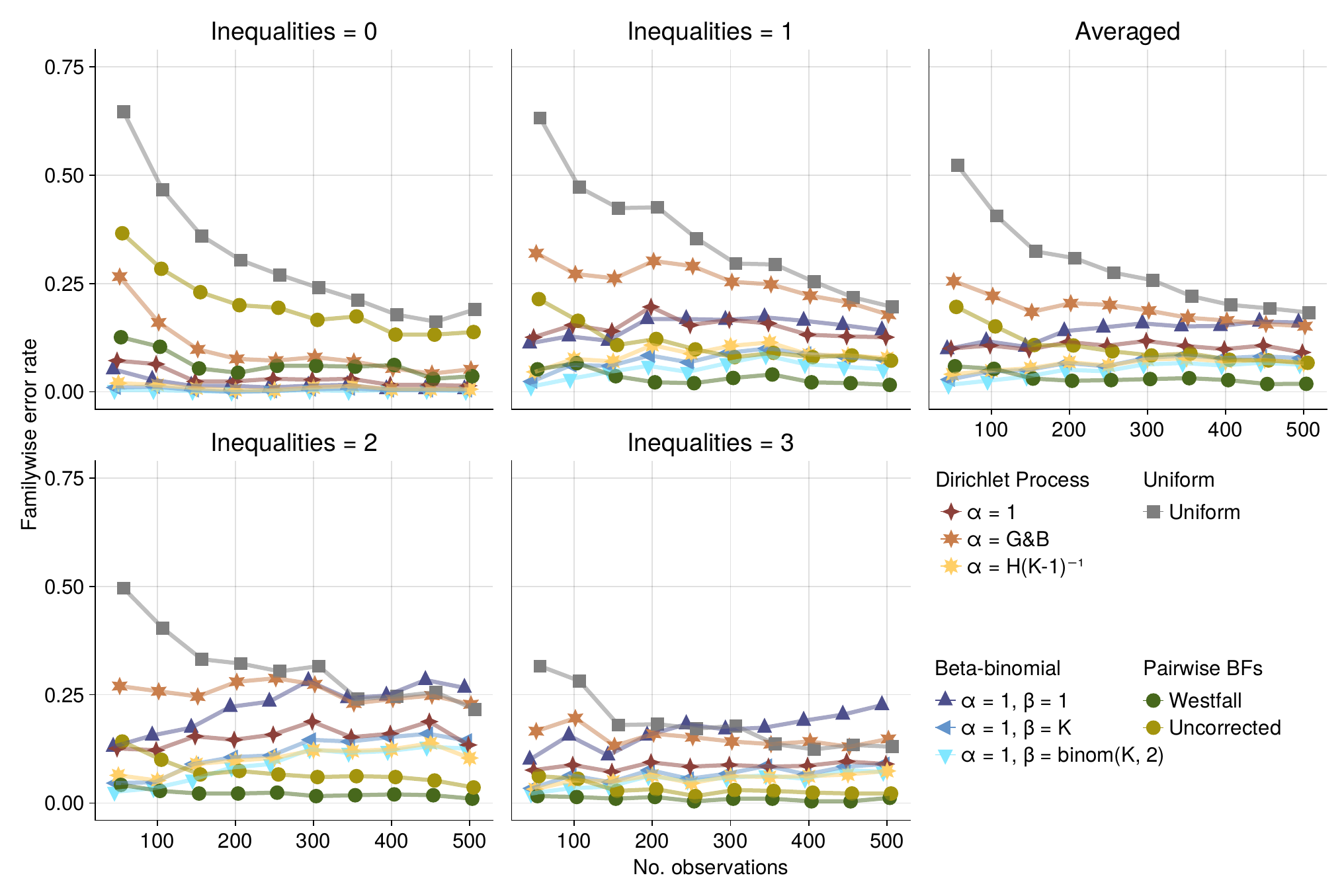}
    \caption{Familywise error rate across priors and sample sizes under a model with 0 (top left), 1 (top right), 2 (bottom left), and 3 (bottom right) true inequalities for $K = 5$ groups. The rightmost panel shows the average familywise error rate across inequalities. 
    }
    \label{fig:big_simulation-I}
\end{figure}

\begin{revision}
Figure~\ref{fig:big_simulation-I} shows the probability of at least one error for different methods across the number of \textit{inequalities} in the true model and sample sizes. The top left panel shows that the uniform prior and the \reviseAS{uncorrected} pairwise Bayes factors perform worst, followed by \reviseAStwo{$\text{DP}(\alpha = \text{G\&B})$} and the method proposed by \textcite{westfall1997bayesian} perform worst. The other Dirichlet process \reviseAS{except $\alpha = 1$} and beta-binomial priors \reviseAS{seem to provide} adequate error control. This mirrors the results above, which is natural since this part of the simulation is a special case for $K = 5$. Increasing the number of inequalities to 1 (\reviseAS{top middle}) and 2 (bottom left), we find that $\text{DP}(\alpha = \text{G\&B})$, $\text{DP}(\alpha = 1)$, and $\text{BB}(\alpha = 1, \beta = 1)$ lead to markedly increased familywise error rates. All other priors lead to increased error rates as well, except the two pairwise Bayes factor methods, whose performance improves. Performance across priors tends to improve again with three inequalities. This may be due to the fact that, with more inequalities, there are simply less opportunities to incorrectly claim that two population means are different.
\end{revision}

\begin{revision}
The rightmost panel in Figure~\ref{fig:big_simulation-I} shows the results averaged over the number of inequalities in the true model. We find that the method by \textcite{westfall1997bayesian} shows the strongest familywise error control, closely followed by the \reviseAStwo{$\text{BB}(\alpha = 1, \beta = K)$} and \reviseAStwo{$\text{BB}(\alpha = 1, \beta = \binom{K}{2})$} and the \reviseAStwo{$\text{DP}(\alpha = \nicefrac{1}{H_{K-1}})$}. The uncorrected pairwise Bayes factor method performs similar to \reviseAStwo{$\text{DP}(\alpha = 1)$}, \reviseAStwo{$\text{BB}(\alpha = 1, \beta = 1)$}, and \reviseAStwo{$\text{DP}(\alpha = \text{G\&B})$}, with the uniform prior performing worst.
\end{revision}

\subsubsection{Statistical Power}
\begin{figure}[!ht]
    \centering
    \includegraphics[width=1\textwidth]{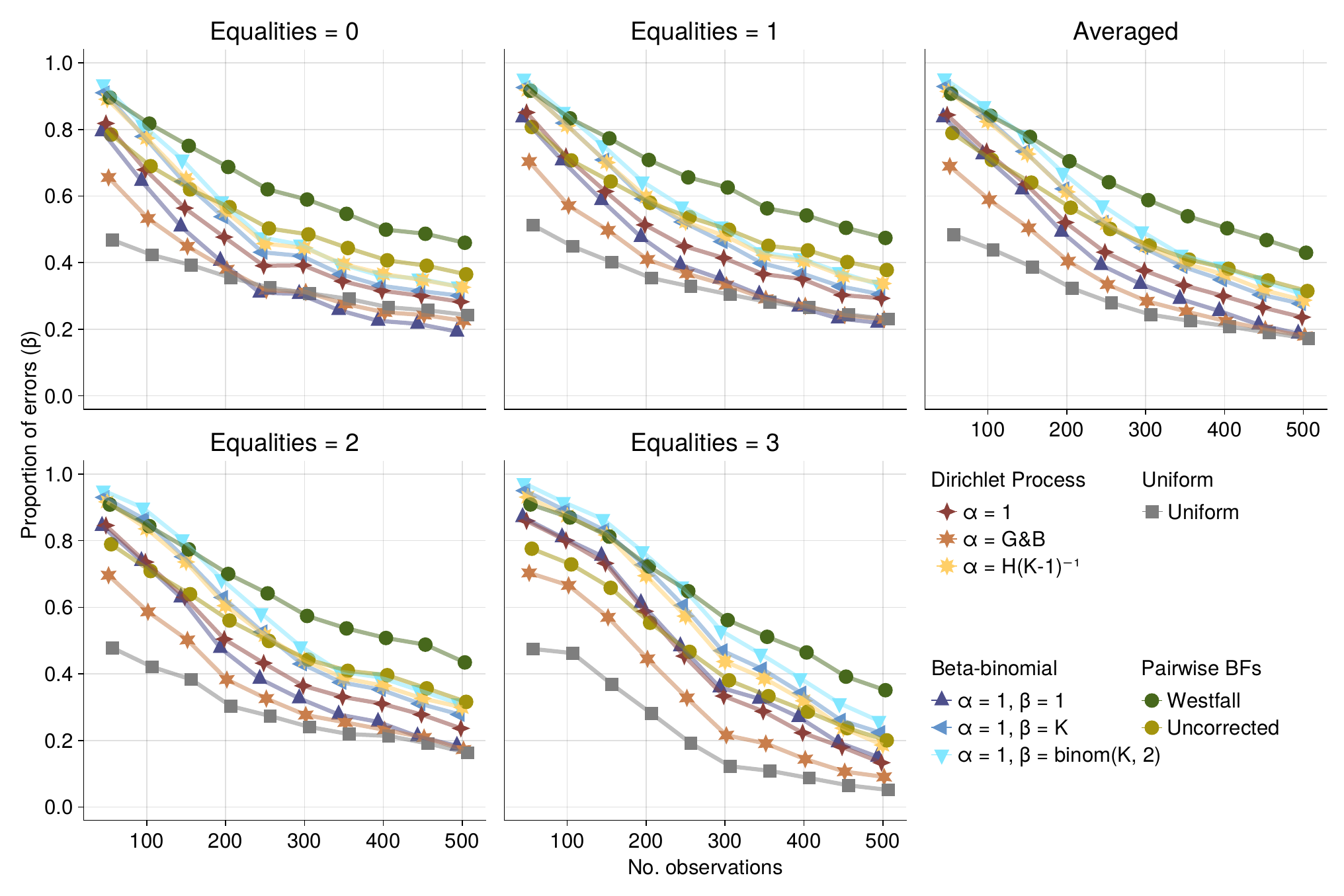}
    \caption{Proportion of falsely claiming a difference between two groups when there is none across priors and sample sizes under a model with 0 (top left), 1 (top right), 2 (bottom left), and 3 (bottom right) true inequalities for $K = 5$ groups. The rightmost panel shows the average error rate across inequalities.}
    \label{fig:big_simulation-II}
\end{figure}

\begin{revision}
Figure~\ref{fig:big_simulation-II} shows the proportion of falsely claiming a difference between two groups when there is none for different methods across the number of \textit{equalities} in the true model and sample sizes. The top left panel shows that the uniform prior, the $\text{DP}(\alpha = \text{G\&B})$, and the $\text{BB}(\alpha = 1, \beta = 1)$ perform best, while the two pairwise Bayes factor methods and the \reviseAStwo{$\text{DP}(\alpha = \nicefrac{1}{H_{K-1}})$} and \reviseAStwo{$\text{BB}(\alpha = 1, \beta = \binom{K}{2})$} priors perform worst. \reviseAStwo{$\text{BB}(\alpha = 1, \beta = K)$} and the \reviseAStwo{$\text{DP}(\alpha = 1)$} are in-between. This relative pattern generally persists with an increasing number of equalities, except that the uncorrected pairwise Bayes factor seems to improve while $\text{BB}(\alpha = 1, \beta = 1)$ seems to worsen in performance. Increasing the number of equalities in the true model, we find that the performance of virtually all methods increases, especially for large sample sizes.

The rightmost panel in Figure~\ref{fig:big_simulation-II} shows the results averaged over the number of equalities in the true model. We find that the method by \textcite{westfall1997bayesian} is highly conservative, trading off the strong familywise error control with an increase in the proportion of false negatives. Similarly, the priors that performed worst with respect to familywise error control — the uniform, \reviseAStwo{$\text{DP}(\alpha = \text{G\&B})$}, and \reviseAStwo{$\text{BB}(\alpha = 1, \beta = 1)$} — perform best here. The other DP and beta-binomial priors as well as the uncorrected pairwise Bayes factors are somewhere in between those two extremes. The differences between the methods become less pronounced with increasing sample size.
\end{revision}

\subsubsection{Simulation Discussion}
Our results show that no single method dominates all others. While the \reviseAStwo{$\text{BB}(\alpha = 1, \beta = 1)$} prior performed best in our initial simulation study described in Section~\ref{sec:illustration}, including models beyond the null and full model showed that this prior performed considerably worse in those settings. \reviseAS{The method proposed by \textcite{westfall1997bayesian} and the uniform prior yield somewhat extreme opposite results, rendering them unsuitable to be chosen as default priors in practice. Interestingly, the uncorrected pairwise Bayes factor method provides familywise error rates and statistical power in between all other priors. The \reviseAStwo{$\text{BB}(\alpha = 1, \beta = K)$} and the \reviseAStwo{$\text{DP}(\alpha = \nicefrac{1}{H_{K-1}})$} prior performed very similarly (cf. Figure~\ref{fig:prior-comparison}), combining the best familywise error control (barring the choice of $\beta = \binom{K}{2}$) with relatively good statistical power. These may those be good default choices in practice, although researchers may choose different priors depending on a cost benefit of type I versus type II errors.} In the next section, we focus on the \reviseAStwo{$\text{BB}(\alpha = 1, \beta = K)$} prior and apply our method to two examples


\section{Applications} \label{sec:applications}
In this section, we apply the beta-binomial setup to two examples: testing the \reviseAS{equality of proportions and variances.} We have developed a generic Julia package called \textit{EqualitySampler.jl} to allow the user to adjust for multiplicity as proposed in this paper.
The code to reproduce the results is given in Appendix~\refAppCode.

\subsection{Comparing Proportions}\label{sec:app:proportions}
\begin{figure}[!ht]
    \centering
    \includegraphics[width=\textwidth]{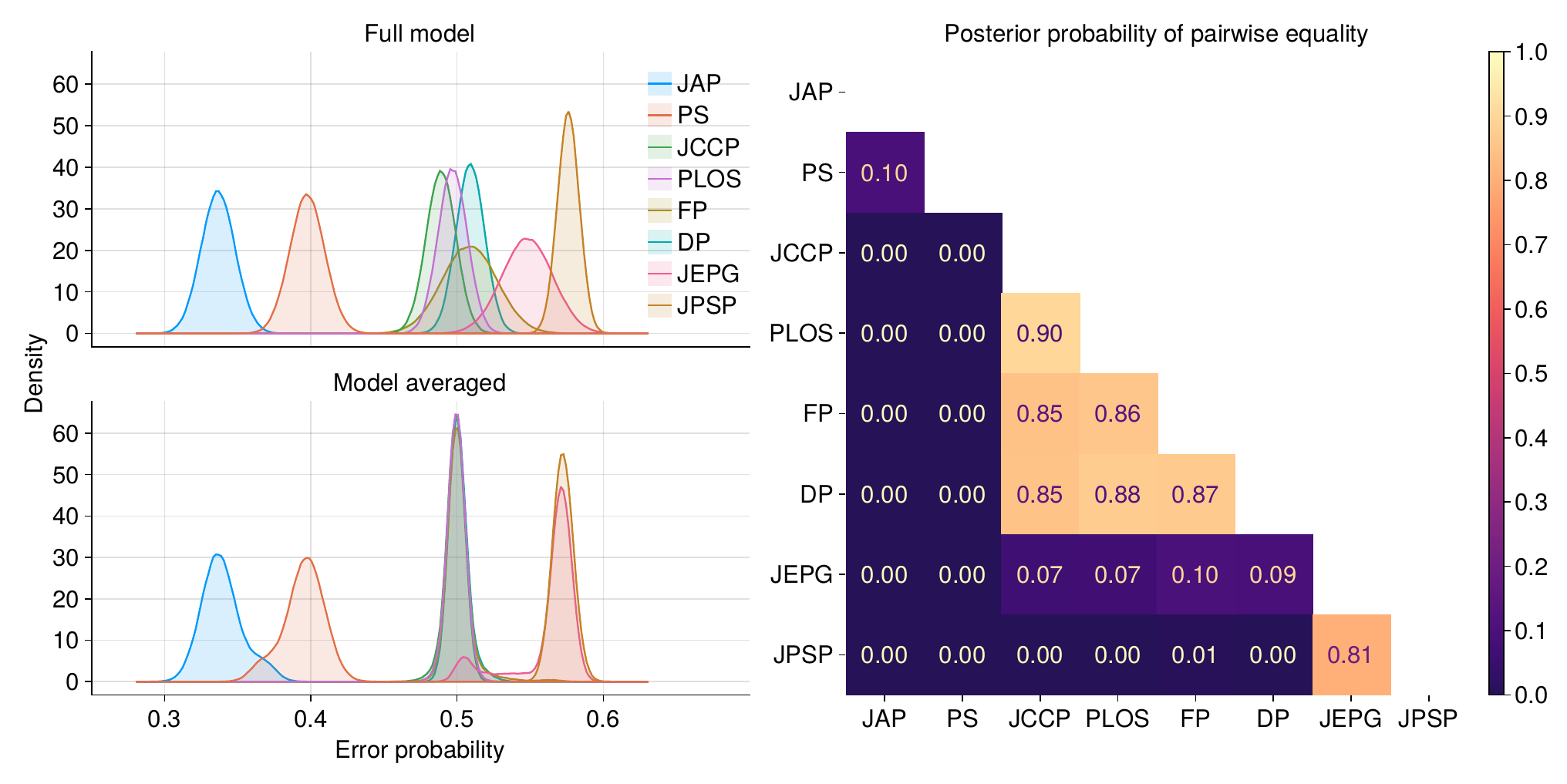}
    \caption{Left: Posterior means of the full model where all proportions are assumed to be different (top) and posterior means when averaging over all models using a beta-binomial($\alpha = 1$, $\beta = 8$) prior (bottom). Right: Posterior probabilities for pairwise equality across all journals. The abbreviations stand for: 
    \emph{Journal of Applied Psychology} (JAP), \emph{Psychological Science} (PS), \emph{Journal of Consulting and Clinical Psychology} (JCCP), \emph{ Public Library of Science} (PLOS), \emph{Developmental Psychology} (DP), \emph{Journal of Experimental Psychology: General} (JEPG), and \emph{Journal of Personality and Social Psychology} (JPSP).}
    \label{fig:demo_proportions}
\end{figure}

\textcite{nuijten2016prevalence} investigated a sample of 30,717 articles published between 1985 and 2013 in eight major psychology journals for statistical reporting errors. Our question here is: Which journals make the same amount of errors, and which make more errors? We answered the question using the following model specification. For journal $j$, denote the number of statistical errors found as $e_j$ and the number of statistical tests analyzed as $n_j$. We assume that underlying each proportion there is a latent true chance of making an error, $\theta_j$. Thus, we model the data as independent binomials, that is, $e_j \sim \mathrm{Binomial}\left(\theta_j, n_j\right)$. Next, we specify a hierarchical level over the partitions to assess for which journals the chances of making an error are equal. This leads to the following model specification:
\begin{align*}
    e_j                 &\sim \mathrm{Binomial}\left(\theta_j, n_j\right)\\
    \theta^u_j          &\sim \text{Beta}(1, 1)\\
    \theta_j            &\leftarrow \theta^c_l\text{ such that } j \in \rho_l\\
    \rho                &\sim \text{beta-binomial}(1, 8) \enspace . \numberthis
\end{align*}
The unconstrained chances $\theta^u_j$ are assigned beta priors from which — together with the partitions — the possibly constrained chances are created. Two chances $\theta_i$ and $\theta_j$ are equal if and only if their indices appear in the same partition $\{i, j\} \subseteq \rho_k$ for some $k$. Note that the model reduces to the full model of independent binomials whenever the partitions state that all elements in $\vec{\theta}$ are distinct. We used a beta-binomial prior with $\alpha = 1$ and $\beta = 8$. The top left panel in Figure~\ref{fig:demo_proportions} shows the posterior distributions for the underlying error chance for each journal under a model that assumes that they are all different.

We can see that the posterior distributions for JCCP (green), PLOS (purple), DP (turquoise), and FP (terracotta) are very close to each other, with FP showing more pronounced uncertainty. The panel below shows the model-averaged posterior distributions, clearly demonstrating a shrinkage effect. The error chances for JAP and PS are pulled toward each other, with JCCP, PLOS, DP, and FP being shrunk towards each other almost completely, similarly to JEPG and JPSP. The right panel in Figure~\ref{fig:demo_proportions} gives the posterior distributions for pairwise equality across all journals, reflecting the two main clusters in the model-averaged density plot on the left. \reviseAS{Note that the frequentist approach to multiple comparisons does not yield such a coherent analysis outcome. Instead, we would first test for overall equality of the error proportions across all journals, and then if this is rejected, engage in pairwise post-hoc comparisons. Adjusting the p-values for multiple comparisons, we would then be able to say, for each of the 28 comparisons, whether the null hypothesis of no difference was rejected. In contrast to our Bayesian approach, the results of this cannot be straightforwardly aggregated into statements of equality and inequality across the groups overall.} \reviseAS{For an example comparing two standard deviations, see Appendix~\refAppStdev. In the next section, we provide an example with much larger $K$.}

\begin{revision}
\subsection{Comparing Means} \label{sec:app:means}
The Program for International Student Assessment (PISA) measures the academic skill of 15-year-old pupils in schools worldwide on subjects such as mathematics and science \parencite{PISA2022VolumeI}. The reports by PISA are used to assess the relative performance of pupils in different countries. Here, we reanalyzed the mathematics scores to assess which countries performed equally well. We obtained the mean score, standard deviation, and sample size for each of the 78 countries \parencite[obtained from Tables I.A2.1. and I.B1.2.1. of][]{PISA2022VolumeI}. We used the same ANOVA model from Section~\ref{sec:illustration} to analyze the data with a $\text{BB}(1, 78)$ prior on the partitions.

While it is best to interpret the model-averaged results whenever possible, sometimes it is desirable to obtain a single partition. For example, one might want to show one final ranking of the countries.  To obtain a single partition, we used the median probability model (MPM).  We obtained the MPM by first exploring the model space as usual, obtaining posterior probabilities for the pairwise inequality of each pair of countries. Next, we thresholded the posterior inequality probabilities using the prior inequality probability. This results in an adjacency matrix $A$ where $A_{ij} = 1$ if $p(\theta_i = \theta_j \mid \mathrm{data}) > p(\theta_i = \theta_j)$ and $A_{ij} = 0$ otherwise. Next, we used a greedy search to find the partition whose equality constraints are closest to $A$. Note that a single optimal partition need not exist, see Appendix~\refAppMpm.

\begin{figure}[!ht]
    \centering
    \includegraphics[width=\linewidth]{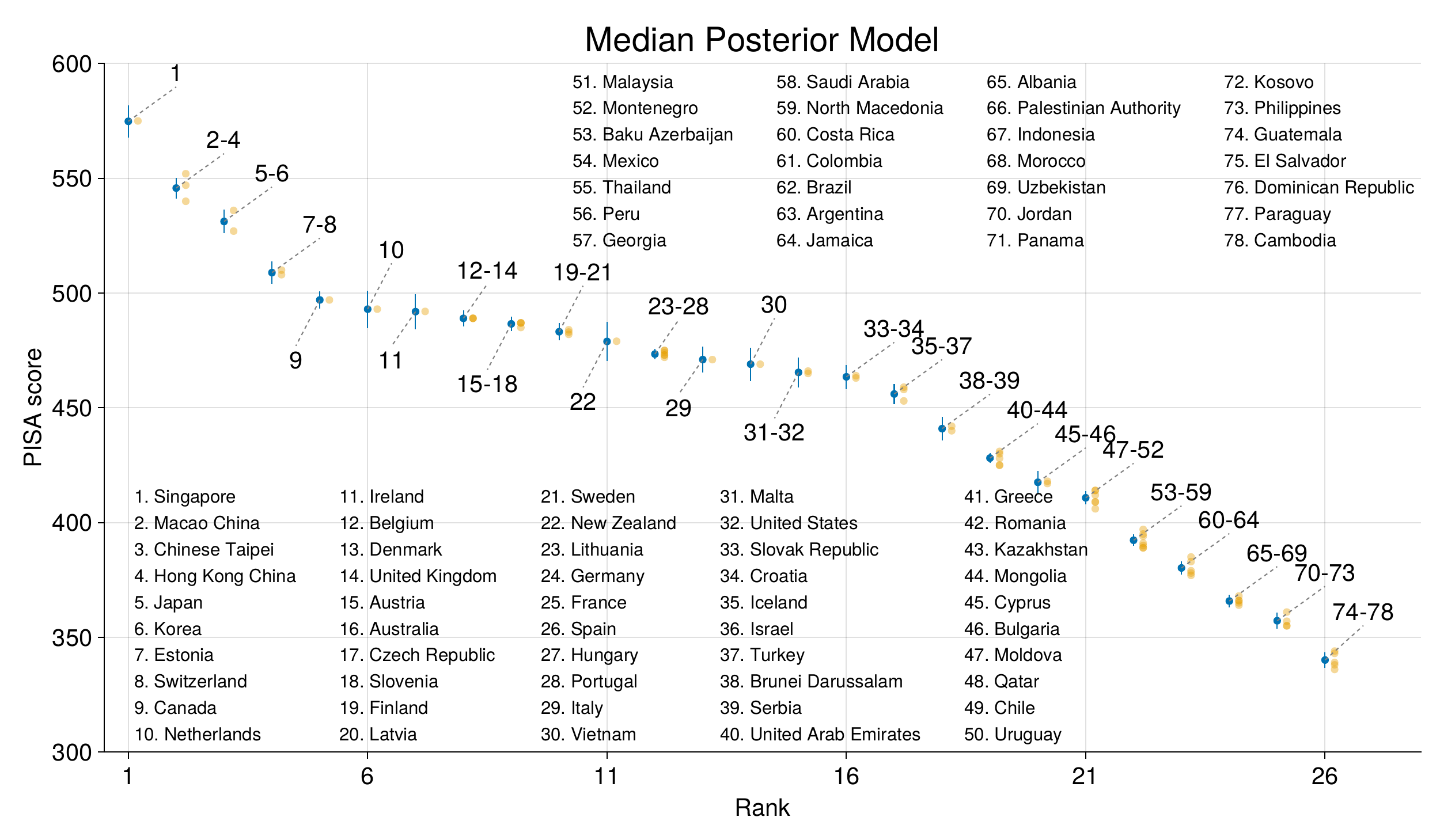}
    \caption{\reviseAS{Results for the median posterior model. The posterior mean and 95\% credible interval are shown in blue for each subset of the partition. 
    The orange dots next to the partition estimate show the sample means. 
    }}
    \label{fig:pisa}
\end{figure}

We found that the order of the countries by posterior mean is identical to the results in the original study \parencite[Table I.2.1][]{PISA2022VolumeI}, see Figure~\ref{fig:pisa}. There are some notable differences between our results and those reported by PISA, however. Specifically, the original authors list, for each country, all other countries that do not differ in a statistically significant manner from the reference country. This leads to non-transitive statements such as, ``the average score between Hong Kong and Japan does not differ significantly, the average score between Japan and Korea does not differ significantly; however, the average score between Hong Kong and Korea does differ significantly.'' In contrast, our approach respects transitivity. For example, our analysis suggests that Japan and Korea have an equal mean score and that the mean score of Hong Kong was larger. Another meaningful difference is that our approach distinguishes many sub-clusters between 10-17 (Netherlands to Czech Republic), whereas \textcite{PISA2022VolumeI} found that these all did not significantly differ. Note that the analysis by \textcite{PISA2022VolumeI} used finite-population corrections, whereas we analyzed the unadjusted data.

\end{revision}

\section{Discussion} \label{sec:discussion}

Testing \reviseAS{equality constraints among} groups while adjusting for multiple comparisons is a core challenge in many applied settings. In this paper, we have proposed a flexible class of beta-binomial priors to penalize multiplicity and make inferences over all possible \reviseAS{equality constraints} in relatively general settings. We compared the beta-binomial priors to a Dirichlet process prior suggested by \textcite{gopalan1998bayesian}, to a uniform prior, to the method proposed by \textcite{westfall1997bayesian}, and to an uncorrected method based on pairwise Bayes factors. \reviseAS{We also established that the Pitman-Yor process, which is a natural extension to the Dirichlet process, provides little benefit in terms of error control or power over the Dirichlet process}. We illustrated our method, which is freely available in the Julia package \textit{EqualitySampler}, on \reviseAS{three} examples.

We found that a beta-binomial prior with $\alpha = 1$ and $\beta \in \{K, \binom{K}{2}\}$ as well as a Dirichlet process prior with $\alpha = \nicefrac{1}{H_{K-1}}$ adequately control the familywise error rate, while the \reviseAS{Dirichlet process} with the specification proposed by \textcite{gopalan1998bayesian}, the uniform prior, and uncorrected pairwise Bayes factors do not. We also found that the method proposed by \textcite{westfall1997bayesian} compares favorably in terms of error control but not in terms of statistical power. While we have focused on a posterior probability threshold of $0.50$ (i.e., a Bayes factor of 1), other thresholds will naturally impact the trade-off between the two types of errors. Importantly, and in contrast to conventional adjustments for multiple comparisons \parencite[e.g.,][]{westfall1997bayesian, jeffreys1961theory}, specifying a prior over the partitions allows inferences over all possible \reviseAS{equality constraints}. This means that researchers can use the methods we provide to assess not only the probability of pairwise \reviseAS{equality constraints} --- as is common in standard post-hoc tests for, say, ANOVA --- but in fact can make probabilistic statements over any set of \reviseAS{equality constraints} they wish to assess. \reviseAS{As illustrated by our reanalysis of PISA data, in contrast to frequentist multiple comparison adjustment methods that rely on post-hoc tests, the Bayesian method outlined in this paper can yield an informative clustering of population means that respects transitivity.} Similarly, the outlined approach also allows for model-averaging, which as we have seen in the applications yields shrinkage of the groups towards each other. \reviseAStwo{Future research may wish to study the effects of assigning a prior to the hyperparameters of the Dirichlet process and beta-binomial priors \parencite[cf.][]{miller2018mixture, ascolani2023clustering}}. 

\revise{%
Ideally, inferences are drawn using the model-averaged posterior distribution. However, sometimes there is a need to select a single model among all candidate models. An obvious choice is the highest posterior density model (HPM). However, as the number of groups increases, it becomes increasingly less likely that a stochastic search algorithm visits the HPM and the sampling uncertainty for the posterior probability of individual models also increases. An alternative that is often used in regression is the median probability model (MPM), which is obtained by retaining all predictors with a posterior probability larger than \nicefrac{1}{2} \parencite{barbieri2004optimal, barbieri2021median}. In the context of multiple comparison adjustment, there is the additional constraint that one wants to select a model that is also a valid partition, and the MPM may not necessarily satisfy this constraint. We present an example where the MPM proposes a model that is not a partition and suggests two possible solutions in Appendix~\refAppMpm.
}


As with any statistical method, there are a number of points to keep in mind. While we suggest default values of $\alpha = 1$ and $\beta = K$ for the beta-binomial prior and $\alpha = \nicefrac{1}{H_{K-1}}$ for the DP prior, researchers may wish to use a more informed prior specification. Values for the prior parameters can be elicited by specifying model priors for two out of the following: the prior on the null model, on the full model, or their ratio. \revise{The resulting probabilities for equality among pairs of parameters can then be assessed by the elicitee. As \textcite[][p. 1132]{gopalan1998bayesian} note, if these do not adequately represent the elicitee's beliefs, and if the elicitee cannot make adjustments of the prior on the null model and the full model to resolve this circumstance, then the Dirichlet process prior is not appropriate. The beta-binomial prior is more flexible in this regard, because it has an additional parameter that can be varied to accommodate prior beliefs (see Table~\ref{tab:overview}). If the elicitee's beliefs cannot be accommodated even with two parameters, then in principle it is also possible to substitute the beta-binomial with a categorical distribution that assigns custom probabilities to partitions of the same size\reviseAStwo{, see Appendix~\refAppStdev}. However, if partitions of the same size should be assigned different prior probabilities, then the beta-binomial prior — and the more general categorical distribution — is inappropriate.} Importantly, the beta-binomial prior differs from the DP prior in that it assigns models with the same number of partitions the same prior probability, while the DP prior assigns more mass to the model with the larger cluster. It is not obvious which of the two behaviors is more desirable, and it may well depend on the problem under study. Researchers using the methods we have made available should keep this difference in mind, although the extent to which it matters in practice remains to be seen.


There are some practical limitations of our implementation that we leave for future work. We currently do not allow for factorial designs, for example, for which dummy or contrast coding is more natural. The key challenge there is to specify the prior in such a way that it reflects the structure of the experimental design. For the present, we believe that the Bayesian approach outlined in this paper can help \revise{improve the inferences of} applied researchers who wish to compare multiple groups.

\if0\blind

\paragraph{Author Contributions.} DvdB and FD proposed the study and refined it in numerous discussions. DvdB implemented the method\reviseAStwo{, created the Julia package, and provided all proofs}. DvdB and FD designed the simulation study. DvdB conducted the simulation study and analyzed the data with the help of FD. DvdB and FD wrote the manuscript. All authors read and approved the submitted version of the paper. They also declare that there were no conflicts of interest.
\fi

\ifTAS
\bigskip
\begin{center}
{\large\bf SUPPLEMENTARY MATERIAL}
\end{center}

\begin{description}



\item[Proofs:] Conditions for which the Dirichlet process prior and beta-binomial prior have decreasing prior model odds, and proof that the predictions rules lead to the desired joint distributions. (pdf) 
\item[Comparison Between the Dirichlet and Pitman-Yor Process.] Comparison with the Pitman-Yor process, a generalization of the Dirichlet process. (pdf) 
\item[Additional Simulation Results:] Simulation results for the conditions where $K = 9$. (pdf)
\item[Selecting the Optimal Model:] Additional discussion on selecting a single optimal model. (pdf)
\item[Example Code:] Example Julia code for the analysis in section~\ref{sec:app:proportions}. (pdf)
\item[Comparing Standard Deviations:] Additional example analysis comparing standard deviations. (pdf)

\end{description}
\fi

\printbibliography

\newpage
\appendix

\ifTAS
\else


\begin{revision}   
\section{Proofs}
\subsection{Dirichlet Process Prior with Decreasing Prior Model Odds} \label{ap:decreasing-odds-dpp}
\begin{prop}\label{prop:monotonicity-dpp}
\reviseAStwo{The prior density on the number of clusters implied by a Dirichlet process over the partitions is decreasing if $\alpha \leq \nicefrac{1}{H_{K-1}}$.}
\end{prop}
\begin{proof}
Given a Dirichlet process prior for $K$ variables with parameter $\alpha$, the probability of \reviseAStwo{a number of clusters} is given by
\begin{align*}
    P(|\partition|\,\mid\, K, \alpha) = 
    \stirlingone{K}{|\partition|} K! \alpha^K 
    \frac{
        \Gamma(\alpha)
        }{
        \Gamma(\alpha + K)
    },
\end{align*}
where $\stirlingone{K}{|\partition|}$ denotes the unsigned Stirling numbers of the first kind. 
We want to find $\alpha$ such that the prior on the size of the partitions is decreasing, i.e., 
\begin{align*}
\frac{
P(|\partition|\,\mid\, K, \alpha) 
}{
P(|\partition| + 1\,\mid\, K, \alpha)
} \geq 1.
\end{align*}
Note that the inequality simplifies to:
\begin{align*}
\frac{
\stirlingone{K}{|\partition|}
}{
\stirlingone{K}{|\partition| + 1} 
} \frac{1}{\alpha} \geq 1.
\end{align*}
The goal is to find a single expression for $\alpha$ that satisfies the inequality for all partition sizes. 
Therefore the first step is to find when $\nicefrac{
\stirlingone{K}{|\partition|}
}{
\stirlingone{K}{|\partition| + 1} 
}$ is minimal.
It can be shown that this ratio is minimal when $|\partition| = 1$.
Using the identities $\stirlingone{K}{1} = (K-1)!$ and $\stirlingone{K}{2} = (K-1)! H_{K-1}$ where $H_{n}$ denotes the n\textsuperscript{th} harmonic number, this ratio simplifies and we obtain $\nicefrac{
\stirlingone{K}{1}
}{
\stirlingone{K}{2} 
} = \nicefrac{1}{H_{K-1}}$.
It follows that the prior on the size of partitions is decreasing if $\alpha \leq \nicefrac{1}{H_{K-1}}$.
\end{proof}
\end{revision}

\begin{revisionTwo}
\begin{prop}\label{prop:monotonicity-dpp_partition}
For any partitions $\rho_n$ and $\rho_m$ of $K$ parameters, if $\setsize{\rho_n} > \setsize{\rho_m}$, then the probability of partition $\rho_n$ is smaller than the probability of partition $\rho_m$ if $\alpha \leq \frac{\Gamma(q)^m q^r}{\Gamma(K-n+1)}$ where $q$ and $r$ are the quotient and remainder such that $K = qm + r$.
\end{prop}
\begin{proof}
From Equation~(\refDirichletPrior{}), we have that 
\begin{equation}
    \pi(\rho \mid \alpha) = \frac{\alpha^{|\rho|}\Gamma(\alpha)}{\Gamma(K + \alpha)} \prod_{c \in \rho} \Gamma(|c|) \enspace ,
\end{equation}
Let $\rho_n$ and $\rho_m$ be two partitions such that $n = |\rho_n| > m = |\rho_m|$.
We wish to find $\alpha$ such that $\pi(\rho_n \mid \alpha) < \pi(\rho_m \mid \alpha)$ or $\pi(\rho_n \mid \alpha) / \pi(\rho_m \mid \alpha) < 1$.
This yields:
\begin{align*}
    \frac{
    \frac{\alpha^n\Gamma(\alpha)}{\Gamma(K + \alpha)} \prod_{c \in \rho_n} \Gamma(|c|)
    }{
    \frac{\alpha^m\Gamma(\alpha)}{\Gamma(K + \alpha)} \prod_{c \in \rho_m} \Gamma(|c|)
    }
    &< 1\\
    \alpha^{n - m}\frac{\prod_{c \in \rho_n} \Gamma(|c|)
    }{\prod_{c \in \rho_m} \Gamma(|c|)}
    &< 1\\
\end{align*}
Next, we maximize the probability of $\rho_n$ and minimize that of $\rho_m$.
This is done by considering $\rho_n$ with a giant component, $\{\{\theta_1, \dots, \theta_{K-n+1}\}, \{\theta_{K-n+2}\}, \dots, \{\theta_{K}\}\}$ which yields $\prod_{c \in \rho_n} \Gamma(|c|) = \Gamma(K-n+1)$.
Similarly, consider for $\rho_m$ a structure where all clusters have approximately equal sizes.
To define this partition, let $q$ and $r$ be the quotient and remainder such that $K = qm + r$. Then, we have a partition with $m-r$ clusters of size $q$ and $r$ clusters of size $q+1$, and $\prod_{c \in \rho_m} \Gamma(|c|) = \Gamma(q)^{m-r}\Gamma(q+1)^{r}=\Gamma(q)^m q^r$.
Letting $n = m +1$, we obtain the following condition for $\alpha$
\begin{align*}
    \alpha
    &< \frac{\prod_{c \in \rho_m} \Gamma(|c|)
    }{\prod_{c \in \rho_n} \Gamma(|c|)}
    = \frac{\Gamma(q)^m q^r}{\Gamma(K-n+1)}
\end{align*}
which completes the proof.

\end{proof}
\end{revisionTwo}

\subsection{Beta-binomial Prior with Decreasing Prior Model Odds} \label{ap:decreasing-odds}
\begin{prop}\label{prop:monotonicity}
\reviseAStwo{For any partitions $\rho_n$ and $\rho_m$ of $K$ parameters, if $\setsize{\rho_n}$ > $\setsize{\rho_m}$, the } prior density of the beta-binomial distribution \reviseAStwo{of $\rho_n$ to $\rho_m$} is decreasing for $\alpha = 1$ and $\beta \geq \binom{K}{2}$, and strictly decreasing for $\alpha = 1$ and $\beta > \binom{K}{2}$.

\end{prop}

\begin{proof}
The prior density of the Beta-binomial over partitions is given by:
\begin{align*}
    \pi\left(\rho \mid K, \alpha, \beta\right) = \binom{K - 1}{|\rho| - 1} 
    \frac{\FBeta{|\rho| - 1 + \alpha}{K - |\rho| + \beta}}
    {\FBeta{\alpha}{\beta}\stirling{K}{|\rho|}} \enspace .
\end{align*}
To examine the ratio of two consecutive model sizes we evaluate the ratio of the prior for partitions $\rho$ and $q$ with $|q| = |\rho|+1$:
\begin{align}\label{eq:betabinomratio}
    \frac{
        \pi\left(\rho \mid K,\,\alpha,\,\beta\right)
    }{
        \pi\left(q \mid K,\,\alpha,\,\beta\right)
    }
    &= 
    \frac{\binom{K - 1}{|\rho| - 1}}{\binom{K - 1}{|\rho|}}
    \,
    \frac{
        \FBeta{|\rho| - 1 + \alpha}{K - |\rho| + \beta}
    }{
        \FBeta{|\rho| + \alpha}{K - |\rho| - 1 + \beta}
    }
    \,
    \frac{\stirling{K}{|\rho|+1}}{\stirling{K}{|\rho|}},
    \\
    &= 
    \frac{|\rho|}{K - |\rho|}
    \,
    \frac{\beta +K-|\rho| -1}{\alpha +|\rho| -1}
    \,
    \frac{\stirling{K}{|\rho|+1}}{\stirling{K}{|\rho|}}.
\end{align}
Using the recurrence relation of the Stirling numbers $\stirling{n+1}{k} = k\stirling{n}{k} + \stirling{n}{k-1}$, the ratio $\nicefrac{\stirling{K}{|\rho|+1}}{\stirling{K}{|\rho|}}$ is equivalent to $\nicefrac{\stirling{K+1}{|\rho|+1}}{\stirling{K}{|\rho|+1}} - (|\rho|+1)$.
This ratio of Stirling numbers was studied by \textcite{berg1975some} and their property 2 provides this inequality
\begin{align*}
    \frac{\stirling{K+1}{|\rho|+1}}{\stirling{K}{|\rho|+1}} - |\rho| - 1 
    \geq
    \frac{\stirling{K+1}{|\rho|}}{\stirling{K}{|\rho|}} - |\rho|.
\end{align*}
It follows that the ratio in Equation \eqref{eq:betabinomratio} is maximal at $|\rho| = K-1$ and has value $\binom{K}{2}$.
Next, we fix $\alpha=1$ and solve $\nicefrac{
        \pi\left(K \mid K-1,\,1,\,\beta\right)
    }{
        \pi\left(K \mid K,\,1,\,\beta\right)
    } = 1$ for $\beta$ which yields $\beta = \binom{K}{2}$.
Thus $\beta\geq \binom{K}{2}$ implies $\pi\left(j+1 \mid K,\,1,\,\beta\right)\geq\pi\left(j \mid K,\,1,\,\beta\right)$ (resp. $\beta > \binom{K}{2}$ implies $\pi\left(j+1 \mid K,\,1,\,\beta\right) > \pi\left(j \mid K,\,1,\,\beta\right)$).
\end{proof}

\begin{revisionTwo}
\subsection{Joint PMFs of the Prediction Rules} \label{app:prediction-rule-proofs}
\begin{prop}\label{prop:predicition-rule-unif}
The prediction rule given
\begin{align*}
    \theta_1 &\sim \mathcal{K},\\
    \theta_{j + 1} \mid \theta_1, \ldots, \theta_j
    &\sim \begin{cases}
    \mathcal{K} & \text{with probability } P_{\pi_{U}} \\
    \text{Categorical}\left(\theta_1^{\star}, \ldots, \theta_{\len_j}^{\star} \mid 1, \ldots, 1\right) & \text{with probability } 1 - P_{\pi_{U}} \enspace ,
    \end{cases} \numberthis
\end{align*}
where
\begin{align}
    P_{\pi_{U}}
    &=
    \frac{\rbellnum{K-j}{\len_j + 1}}{\rbellnum{K-j}{\len_j + 1} + \len_j\rbellnum{K-j}{\len_j}}
\end{align}
implies that the distribution over partitions is uniform, i.e., $p(\rho) = \frac{1}{\bellnum{K}}$ where $K = \setsize{\rho}$.
\end{prop}
\begin{proof}
\textcite[p.7]{mezo2011r} show the following recurrence relation for the $r$-Bell numbers:
\begin{align*}\label{prop:mezo7}
    \rbellnum{K}{r} = \rbellnum{K-1}{r+1} + r\rbellnum{K-1}{r}.
\end{align*}
It follows that the denominator of $P_{\pi_{U}}$ may be simplified and we obtain 
\begin{align}
P_{\pi_{U}} = \frac{\rbellnum{K-j}{\len_j + 1}}{\rbellnum{K-j+1}{\len_j}},\quad\text{and}\quad
1 - P_{\pi_{U}} &= \frac{\len_j\rbellnum{K-j}{\len_j}}{\rbellnum{K-j+1}{\len_j}}.
\end{align}
The joint distribution of a partition can be expressed by multiplying the prediction rule for $j \in 2, \dots, K$, omitting the base distribution.
Regardless of whether a new or previously sampled value is drawn, the numerator of the prediction rule for $j$ cancels against the denominator of $j + 1$. If the process is at $K, j, r_j$ and a new value is drawn then $r_{j+1} = r_j + 1$ and $\rbellnum{K-j}{\len_j + 1} = \rbellnum{K-(j+1)+1}{\len_{(j+1)}}$.
The $r$-Bell numbers cancel similarly when a previously observed value is drawn.
It follows that the joint distribution equals
\begin{align*}
    \prior{\rho} &= \prod_{j=2}^K
        \prior{\rho_{j+1:K}\mid \rho_{1:j-1}} 
        = \prod_{j=2}^K \left(P_{j, \pi_{U}}\right)^{\mathbb{I}(\theta_j \notin \theta_1, \dots, \theta_{j-1})} \left(\frac{1 - P_{j, \pi_{U}}}{r_j}\right)^{\mathbb{I}(\theta_j \in \theta_1, \dots, \theta_{j-1})} \\
        &= \frac{1}{\rbellnum{K - 1}{1}} = \frac{1}{\bellnum{K}}.
\end{align*}
Note that the $r$-Bell number in the numerator of $P_{K, \pi_{U}}$ and $1 - P_{j, \pi_{U}}$ equals $\rbellnum{0}{r_K} = 1$, leaving only the denominator of $P_{2, \pi_{U}}$. 
The simplification from the $r$-Bell to Bell numbers can be derived from Equation~\ref{prop:mezo7}.
This proves that the joint distribution is uniform, as desired.
\end{proof}

\begin{prop}\label{prop:predicition-rule-bb}
Instead of proving the result directly for the Beta-binomial, we provide a more general proof where an arbitrary probability vector gives the probability of a particular number of clusters.

Let $\bm{\tilde{p}}$ contains arbitrary probabilities, i.e., $0 \leq \tilde{p}_1, \tilde{p}_2, \dots, \tilde{p}_K$ such that $\sum_{i=1}^K \tilde{p}_i = 1$ and let $p_i = \tilde{p}_i / {\stirling{K}{|\rho|}}$.
Here, $\tilde{p}_i$ equals the total probability of observing $i$ clusters, whereas $\tilde{p}_i$ equals the probability of a specific partition with $i$ clusters.

Consider the following prediction rule:
\begin{align*}
    \theta_1 &\sim \mathcal{K},\\
    \theta_{j + 1} \mid \theta_1, \ldots, \theta_j &\sim \begin{cases}
    \mathcal{K} & \text{with probability } P_{\pi} \\
    \text{Categorical}\left(\theta_1^{\star}, \ldots, \theta_\len^{\star} \mid 1, \ldots, 1\right) & \text{with probability } 1 - P_{\pi} \enspace .
    \end{cases} \enspace , \numberthis
\end{align*}
where
\begin{align}
    P_{j, \pi_S} &= 
    \frac{
        \sum_{i=1}^K p_i \rstirling{K-j+\len+1}{i}{\len + 1}
    }{
        \len \sum_{i=1}^K p_i \rstirling{K-j+\len}{i}{\len} +
             \sum_{i=1}^K p_i \rstirling{K-j+\len+1}{i}{\len + 1}
    } \enspace ,
\end{align}
and $\len_j$ is the number of unique parameters in $\theta_1,\dots,\,\theta_j$.

This prediction rule implies the following joint distribution over partitions that only depends on the $\bm{\tilde{p}}$ and the number of clusters:
\begin{equation}
    \pi(\rho \mid K) = p_{\setsize{\rho}}
\end{equation}

\end{prop}
\begin{proof}
\textcite[p.7]{mezo2011r} lists the following recurrence relation for the $r$-stirling numbers:
\begin{align}
    \rstirling{K+r}{n+r}{r} = \rstirling{K+r}{n+r}{r - 1} - (r - 1)\rstirling{K+r - 1}{n+r}{r - 1}
\end{align}
from which we may simplify the prediction rule to
\begin{align}
    P_{j, \pi_S} &=
    \frac{
        \sum_{i=1}^K p_i \rstirling{K-j+\len_j+1}{i}{\len_j + 1}
    }{  \sum_{i=1}^K p_i \rstirling{K-j+\len_j+1}{i}{\len_j}
    } \\
    1 - P_{j, \pi_S} &=
    \frac{
        r\sum_{i=1}^K p_i \rstirling{K-j+\len_j}{i}{\len_j}
    }{   \sum_{i=1}^K p_i \rstirling{K-j+\len_j+1}{i}{\len_j}
    }
\end{align}
The proof proceeds similarly to that of the uniform. 
The numerator of the prediction rule for $j$ cancels against the denominator of $j + 1$. 
If the process is at $K, j, r_j$ and a new value is drawn then $r_{j+1} = r_j + 1$ and $\rstirling{K-j+\len_j+1}{i}{\len_j + 1} = \rstirling{K-(j+1)+\len_{j+1}+1}{i}{\len_{j+1}}$.
Thus, the joint distribution is given by the numerator for $j = K$ over the denominator for $j = 2$:
\begin{align*}
    \prior{\rho} &= \prod_{j=2}^K
        \prior{\rho_{j+1:K}\mid \rho_{1:j-1}} 
        = \prod_{j=2}^K \left(P_{j, \pi_{U}}\right)^{\mathbb{I}(\theta_j \notin \theta_1, \dots, \theta_{j-1})} \left(\frac{1 - P_{j, \pi_{U}}}{r_j}\right)^{\mathbb{I}(\theta_j \in \theta_1, \dots, \theta_{j-1})} \\
        &= 
        \frac{
            \sum_{i=1}^K p_i\rstirling{r_K}{i}{r_K}
        }{
            \sum_{i=1}^K p_i\rstirling{K}{i}{1}
        } = p_{r_K}
\end{align*}
In the last line, the denominator for $j = 2$ simplifies to 1 as $\rstirling{K-2+1 + 1}{i}{1} = \stirling{K}{i}$.
In the numerator, we have that $\rstirling{r_K}{i}{r_K}$ equals 1 if $i = r_K$ and 0 otherwise.
Since $r_K = \setsize{\rho}$, this completes the proof.
\end{proof}

\end{revisionTwo}

\begin{revision}
\section{Comparison Between the Dirichlet and Pitman-Yor Process}\label{ap:py_vs_dpp}
A natural extension of the Dirichlet is the Pitman-Yor process. The Pitman-Yor process generalizes the Dirichlet process with a discount parameter $d$ with $0\leq d < 1$ and reduces to the Dirichlet Process when $d = 0$. The prediction rule for the Pitman-Yor Process is given by:
\begin{equation} \label{eq:prediction-rule-PY}
    \theta_{j + 1} \mid \theta_1, \ldots, \theta_j\sim \begin{cases}
    \mathcal{K} & \text{with probability } \frac{\alpha + rd}{\alpha+ j - 1} \\
    \text{Categorical}\left(\frac{\theta_1^{\star} - d}{\alpha + j - 1}, \ldots, \frac{\theta_\len^{\star} - d}{\alpha + j - 1} \mid n^{\star}_1, \ldots, n^{\star}_\len\right) & \text{with probability } 1 - \frac{\alpha + rd}{\alpha+ j - 1} \enspace .
    \end{cases} \numberthis
\end{equation}
The key difference is that the discount parameter is a function of the existing number of groups $r$. We observed that the Pitman-Yor Process offers little benefit over the Dirichlet process for the multiple comparison problem, as its behavior can be mimicked by choosing $\alpha$ accordingly. Specifically, we simulated data with $K = 5$, $n = 100$, and varied the true number of equalities to be $\{0\%, 25\%, 50\%, 75\%, 100\%\}$, repeating each configuration 200 times. We analyzed each dataset with multiple Pitman-Yor process and Dirichlet process priors.

The left panel in Figure~\ref{app:fig:py_vs_dpp} shows that the Pitman-Yor process prior obtains the best familywise error control when the discount parameter tends to 0, essentially reducing it to a Dirichlet process. The right panel shows that the Dirichlet process performs worse than the Pitman-Yor in terms of statistical power for the same $\alpha$, but as $\alpha$ parameter increases the Dirichlet process gains in power. 
It appears that the impact of different values for the discount parameters can also be achieved by adjusting $\alpha$ accordingly.
For example, suppose we take the Pitman-Yor prior with $d = \nicefrac{3}{4}$ and $\alpha = 2$. 
Then the probability of at least one error is roughly .7 and the probability of errors is about .2. 
However, had we used a Dirichlet process with $\alpha = 6$ then we would obtained similar results, a type-I error of $\approx .7$ and a $\beta$ of $\approx .2$.
The same goes for e.g., the Pitman-Yor prior with $d = \nicefrac{1}{4}$ and $\alpha = 3$ and a Dirichlet process with $\alpha = 4$, both lead to a type-I error of approximately .45 and a $\beta$ of approximately .38.
In sum, it appears that the parameters $\alpha$ and $d$ play a similar role and boil down to a trade-off between familywise error control and power.
The added complexity of the Pitman-Yor process prior seems to add little in addition to simply varying $\alpha$ in the Dirichlet process prior as a function of $K$.

\begin{figure}
    \centering
    \includegraphics[width=\linewidth]{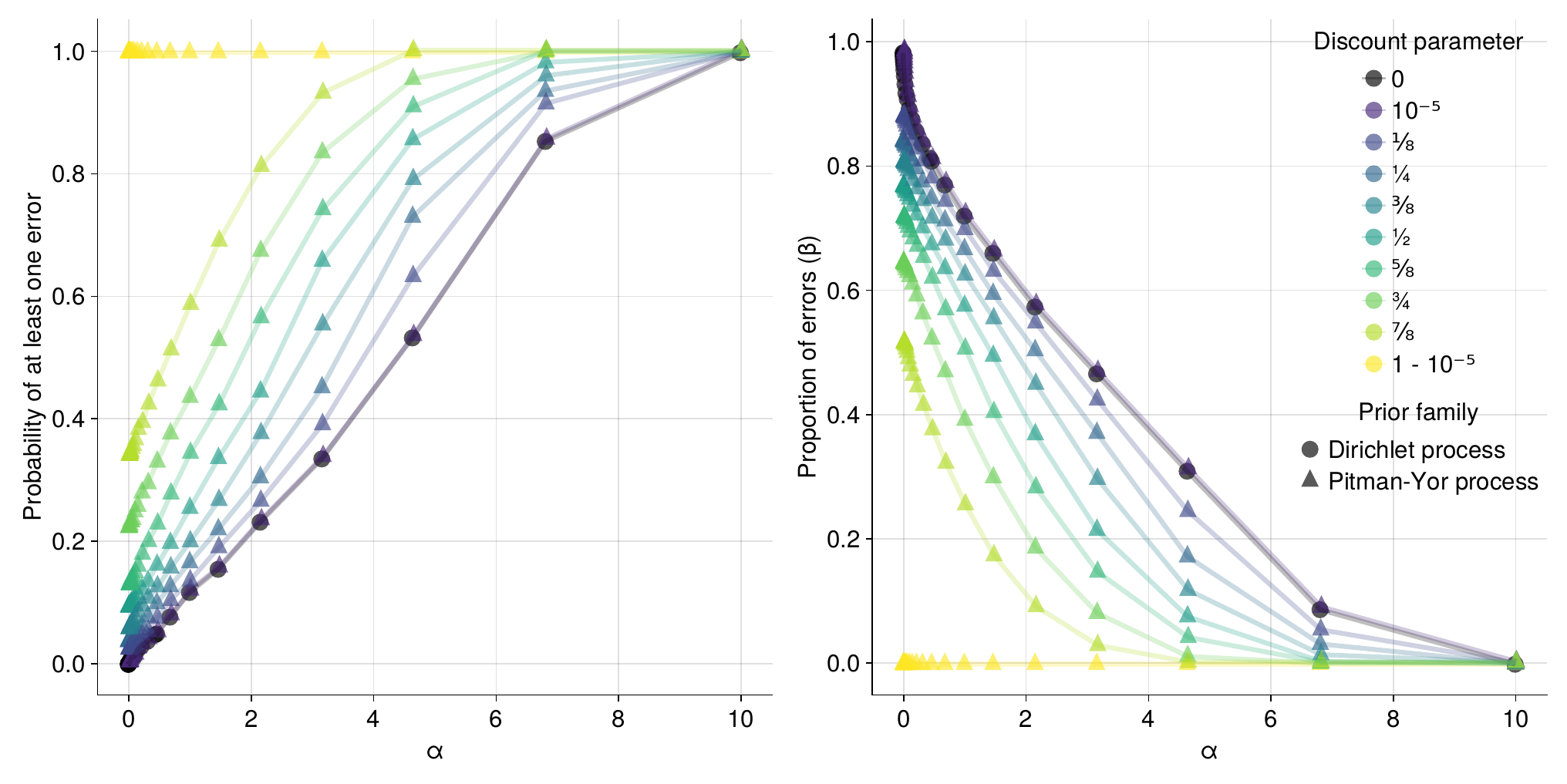}
    \caption{\reviseAS{Comparison between the Pitman-Yor and Dirichlet process priors in terms of familywise error rate (left) and statistical power (right). The Pitman-Yor and Dirichlet process priors coincide for $d = 0$.
    }}
    \label{app:fig:py_vs_dpp}
\end{figure}

\end{revision}

\section{Simulation Results for $K = 9$} \label{app:simulation}
Here we present the extended simulation results for the $K = 9$ group case. Figure~\ref{fig:big_simulation-k9-I} mirrors the results for the $K = 9$ case, namely that the pairwise Bayes factors, the method proposed by \textcite{westfall1997bayesian}, and the uniform prior generally increase in performance as the number of inequalities increase, while the other priors generally decrease in performance. Averaging over the settings, we again find that the beta-binomial prior with $\beta = 1$, the uniform prior, and the symmetric DP prior exhibit the worst error control, with the method proposed by \textcite{westfall1997bayesian} performing best, closely followed by the beta-binomial prior with $\beta = \binom{K}{2}$ and the DP prior with $\alpha = 0.50$.

\begin{figure}[!ht]
    \centering
    \includegraphics[width=1\textwidth]{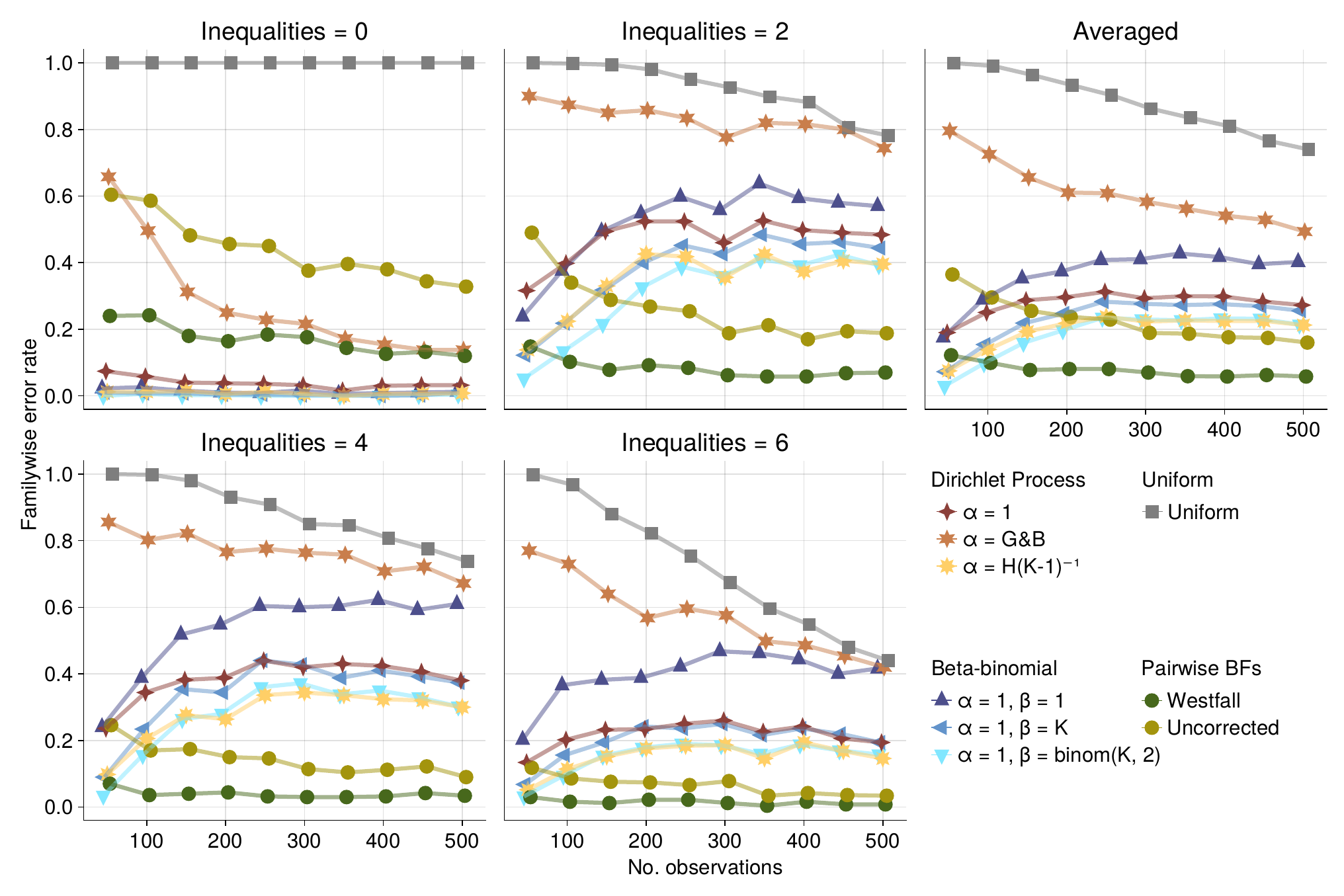}
    \caption{Familywise error rate across priors and sample sizes under a model with 0 (top left), 3 (top right), 5 (bottom left), and 7 (bottom right) true inequalities for $K = 9$ groups. The rightmost panel shows the average familywise error rate across inequalities.}
    \label{fig:big_simulation-k9-I}
\end{figure}

\begin{figure}[!ht]
    \centering
    \includegraphics[width=1\textwidth]{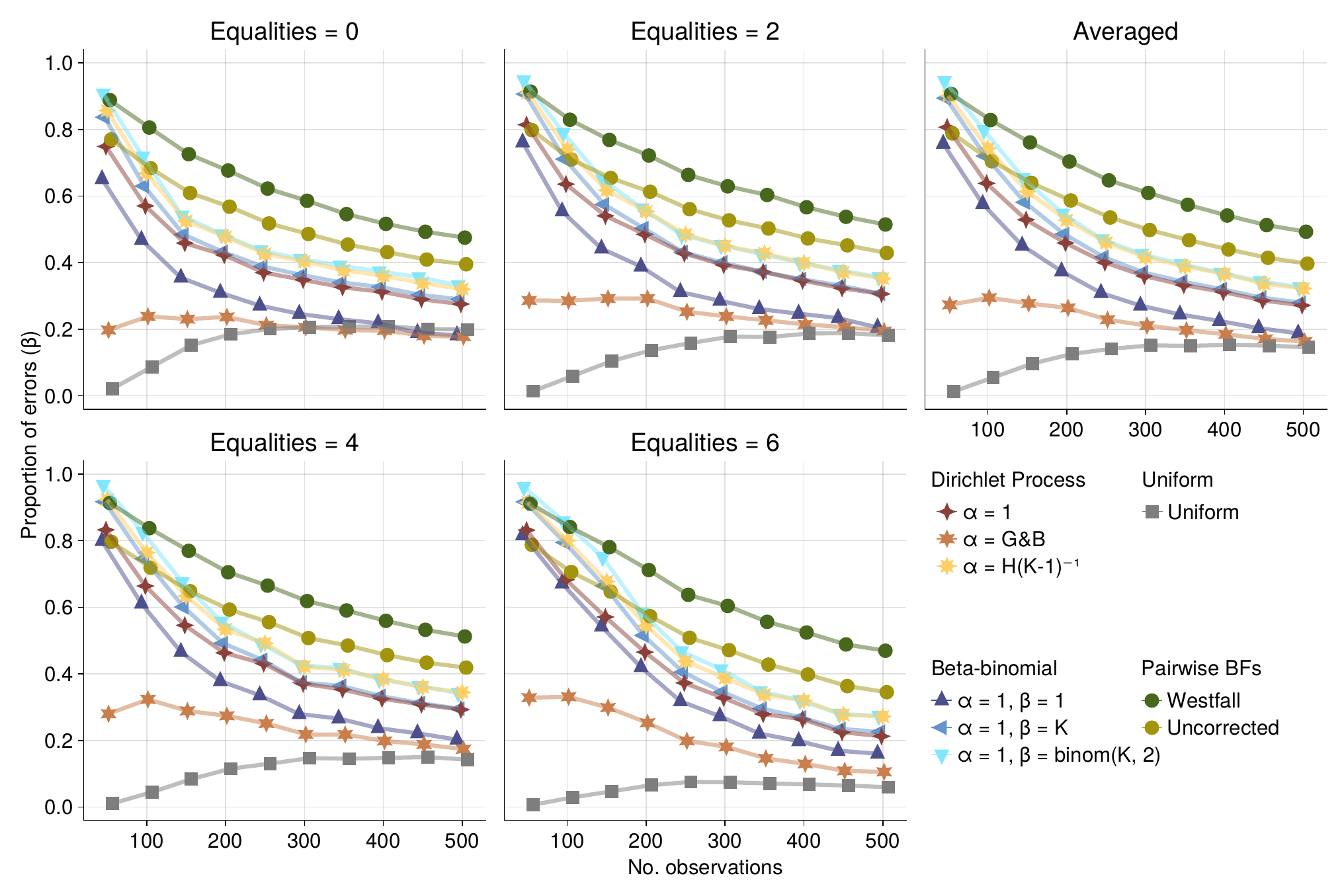}
    \caption{Proportion of falsely claiming a difference between two groups when there is none across priors and sample sizes under a model with 0 (top left), 3 (top right), 5 (bottom left), and 7 (bottom right) true inequalities for $K = 9$ groups. The rightmost panel shows the average error rate across inequalities.}
    \label{fig:big_simulation-k9-II}
\end{figure}

\section{\revise{Selecting the Optimal Model}}\label{ap:single_model}
The median probability model (MPM) is not guaranteed to select a valid partition. To see this, suppose we have obtained an equal number of posterior samples for the partitions $\{\{\theta_1, \theta_2\}, \{\theta_3, \theta_4\}\}$, $\{\{\theta_1\}, \{\theta_2, \theta_3, \theta_4\}\}$, and $\{\{\theta_1, \theta_2, \theta_3\}, \{\theta_4\}\}$. Table~\ref{tab:mpm_example} shows the resulting model-averaged probabilities of equality.

\begin{table}[!ht]
    \caption{An example of model-averaged posterior probabilities where the median probability model is not a valid partition.}
    \label{tab:mpm_example}
    \centering
    \begin{tabular}{lccc}
        \toprule
                  & $\theta_1$      & $\theta_2$      & $\theta_3$     \\
        \midrule
        $\theta_2$& \nicefrac{2}{3} &                 &                \\
        $\theta_3$& \nicefrac{1}{3} &  \nicefrac{2}{3}&                \\
        $\theta_4$& \nicefrac{0}{3} &  \nicefrac{1}{3}& \nicefrac{2}{3}\\
         \bottomrule
    \end{tabular}
\end{table}

The posterior probabilities larger than \nicefrac{1}{2} are $p(\theta_1 = \theta_2) = \nicefrac{2}{3}$, $p(\theta_2 = \theta_3) = \nicefrac{2}{3}$, and $p(\theta_3 = \theta_4) = \nicefrac{2}{3}$. Transitivity would imply that $\theta_1 = \theta_4$, but we have that $p(\theta_1 = \theta_4) = \nicefrac{0}{3} < \nicefrac{1}{2}$. The MPM therefore does not yield a valid partition. 

We propose two solutions that are inspired by the median probability model.
The first solution is to find a partition that is closest to the model-averaged posterior probabilities of equality. That is, we want a partition $\rho$ that minimizes

\begin{equation}
\sum_{i<j}\left( I(\rho \shortimplies \theta_i = \theta_j) - p(\theta_i = \theta_j \mid \vec{y})\right)^2 \enspace ,
\end{equation}
where $I(\rho \shortimplies \theta_i = \theta_j)$ is 1 if the partition $\rho$ implies that $\theta_i$ equals $\theta_j$ and 0 otherwise. For the example in Table~\ref{tab:mpm_example}, this solution suggests that the optimal partition is $\{\{\theta_1, \theta_2\}, \{\theta_3, \theta_4\}\}$. The second approach is similar to the first, but rather than minimizing the distance to the probabilities of equality, we minimize the distance to the model-averaged posterior distributions on the level of the parameters. We do so by finding a partition $\rho$ that minimizes $d(p(\theta\mid \vec{y}, \rho), p(\theta \mid \vec{y}))$, where $d$ is a distance function, $p(\theta\mid \vec{y}, \rho)$ is the posterior distribution conditional on partition $\rho$, and $p(\theta \mid \vec{y})$ is the model averaged posterior. In principle, any distance function can be used, but in regression contexts it is common to use the squared distance between the mean of the distributions. The first solution is a discrete optimization problem that is relatively easy to carry out using existing software for integer programming \parencite[e.g., JuMP;][]{lubin2023jump}. The second solution is more complex to carry out because it may be necessary to resample from the posterior distribution for a given partition, but also closer to the optimality condition for the MPM of \textcite[][Eq. 16]{barbieri2004optimal}.

\section{Example Code} \label{sec:appendix-code}
The code below illustrates the proportion example in Section~\refSecProps.
To install the package enter the Pkg REPL by typing \lstinline[language=Julia]{]} and \lstinline[language=Julia]{add EqualitySampler}.
Alternatively, the package can be installed by importing the Pkg package: \lstinline[language=Julia]{import Pkg; Pkg.add("EqualitySampler")}.

\begin{lstlisting}
using EqualitySampler
import DataFrames    as DF,
     LinearAlgebra   as LA,
     NamedArrays     as NA,
     CSV

# working directory is assumed to be the root of the GitHub repository
journal_data = DF.DataFrame(CSV.File(joinpath("simulations", "demos", "data",
  "journal_data.csv")))

# K
n_journals = size(journal_data, 1)

# no of observed errors
errors = round.(Int, journal_data.n .* journal_data.errors)
# no of possible errors
observations = journal_data.n

# no. MCMC iterations
no_iter = 200_000

# no. groups, i.e., K
no_journals = length(journal_data.journal)

partition_prior = BetaBinomialPartitionDistribution(no_journals, 1, no_journals)

# with EqualitySampler.EnumerateThenSample first enumerates the model space
# and then resamples to obtain model-averaged parameter distributions.
prop_samples_eq   = proportion_test(total_counts, no_errors, 
  EqualitySampler.EnumerateThenSample(iter = no_iter), partition_prior)

# compute the posterior probability of equality
eq_prop_mat = compute_post_prob_eq(prop_samples_eq)

# The posterior probability that two journals are equal
NA.NamedArray(
  LA.UnitLowerTriangular(round.(eq_prop_mat; digits = 2)),
  (journal_data.journal, journal_data.journal)
)
8x8 Named LinearAlgebra.UnitLowerTriangular{Float64, Matrix{Float64}}
A \ B |  JAP    PS  JCCP  PLOS    FP    DP  JEPG  JPSP
------|-----------------------------------------------
JAP   |  1.0   0.0   0.0   0.0   0.0   0.0   0.0   0.0
PS    |  0.1   1.0   0.0   0.0   0.0   0.0   0.0   0.0
JCCP  |  0.0   0.0   1.0   0.0   0.0   0.0   0.0   0.0
PLOS  |  0.0   0.0   0.9   1.0   0.0   0.0   0.0   0.0
FP    |  0.0   0.0  0.85  0.86   1.0   0.0   0.0   0.0
DP    |  0.0   0.0  0.85  0.88  0.87   1.0   0.0   0.0
JEPG  |  0.0   0.0  0.07  0.07   0.1  0.09   1.0   0.0
JPSP  |  0.0   0.0   0.0   0.0  0.01   0.0  0.81   1.0
# The table above is identical to the right panel of Figure 7.
\end{lstlisting}

\section{\reviseAS{Comparing} Standard Deviations} \label{app:standard-deviation}

\textcite{borkenau2013sex} studied whether men and women differ in the variability of personality traits. Here we focus on five personality traits (agreeableness, extraversion, openness, conscientiousness, neuroticism) rated by participants' peers in an Estonian sample consisting of $n_1 = 969$ women and $n_2 = 716$ men. Our goal is to assess which personality traits across the sexes can be assumed equal in terms of their variability. This example shows how our methodology can be used to test group differences while taking the multivariate dependency of the outcome measure into account. We build on the parameterization proposed by \textcite{dablander2020default}, who developed a default Bayes factor test for testing the (in)equality of variances. Let $\vec{y}_1$ and $\vec{y}_2$ denote the five-element vectors of observed data for men and women, respectively, and $K = 10$ be the total number of variables. For each sex $k \in \{1, 2\}$, we have:
\begin{align*}
    \vec{Y}_k         &\sim \mathcal{N}\left(\vec{\mu}_k, \Sigma_k \right)\\
    \vec{\mu}_k       &\propto \vec{1}  \\
    \Sigma_k          &= \text{diag}(\vec{\sigma}_k) \, \Omega_k \, \text{diag}(\vec{\sigma}_k) \\
    \Omega_k          &\sim \text{LKJ}(1) \enspace .
\end{align*}
\reviseAS{where $\mu \propto 1$ indicates an improper prior and $\sigma^2 \propto 1 / \sigma^2$ indicates Jeffreys's prior, both of which are routinely applied for testing purposes \parencite[e.g.,][]{ly2016harold}. The Lewandowski-Kurowicka-Joe (LKJ) prior has its hyperparameter set to 1, implying a uniform prior over the space of correlation matrices.}
To test the equality of variances both between and across groups, we define the ten-variable standard deviation vector $\vec{\sigma} = [\vec{\sigma}_1, \vec{\sigma}_2]$ with $\bar{\sigma}$ denoting the average standard deviation. Following \textcite{dablander2020default}, we write $\sigma_j = \left(K\vartheta_j\bar{\sigma}\right)^{-1}$, where $\vartheta_j = \nicefrac{\sigma_j}{\sum_{j = 1}^K \sigma_j}$ is the relative standard deviation and $\vartheta_K = 1 - \sum_{j = 1}^{K - 1} \vartheta_j$. To complete the model specification, we write:
\begin{align*}
    \sigma_j     &= \left(K\vartheta_j\bar{\sigma_j}\right)^{-1} \\
    \bar{\sigma}_j &\propto \bar{\sigma}_j^{-1} \\
    \vartheta_j            &\leftarrow \text{mean of elements of } \vartheta^u \text{ in the same partition }\\
    \vec{\vartheta}^u     &\sim \text{Dirichlet}(1, \ldots, 1) \\
    \rho              &\sim \text{beta-binomial}(1, 10) \enspace . \numberthis
\end{align*}
Two standard deviations $\sigma_i$ and $\sigma_j$ are equal if and only if their indices appear in the same partition $\{i,j\}\subseteq\partition_k$ for some $k$. When the partition states that all standard deviations are distinct we recover the full model. 
The top left panel of Figure~\ref{fig:demo_variances} shows the posterior distributions under the full model that assumes all standard deviations are different.

\begin{figure}[!ht]
    \centering
    \includegraphics[width=\textwidth]{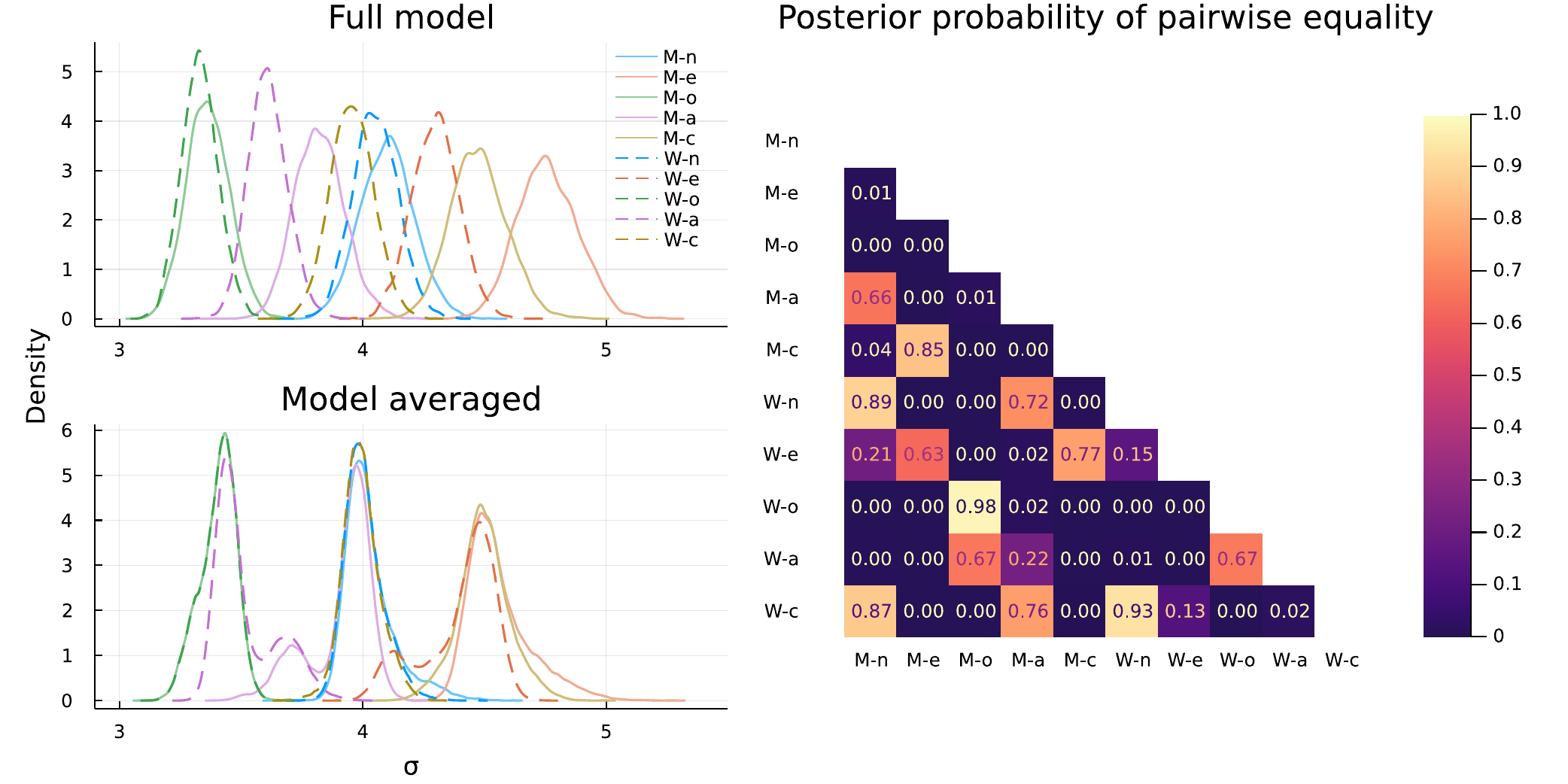}
    \caption{Left: Posterior means of the full model where all standard deviations are assumed to be different (top) and posterior means when averaging across all models using a beta-binomial($\alpha$ = 1, $\beta$ = 10) prior (bottom).
    Right: Posterior probabilities for pairwise equality across all personality traits. In the abbreviations the first letter stands for \emph{men} (m) or \emph{women} (w). The second letter stands for \emph{neuroticism} (n), \emph{extraversion} (e), \emph{openness} (o), \emph{agreeableness} (a), and \emph{conscientiousness} (c).}
    \label{fig:demo_variances}
\end{figure}

While all posterior distributions lie close to each other, the standard deviations of openness for men and women overlap particularly strongly. The bottom panel shows the model-averaged posterior distributions, which again demonstrate a shrinkage effect. The right panel of Figure~\ref{fig:demo_variances} shows the posterior probability of pairwise equality across all personality traits for men and women. It appears that there are three clusters: (1) men--openness, women--openness, and women--agreeableness; (2) men--neuroticism, women--neuroticism, women--conscientiousness, and men--agreeableness; (3) men--conscientiousness, men--extraversion, and women--extraversion. However, for the personality traits women--agreeableness, men--agreeableness, and women--extraversion, the evidence is not overwhelming, as indicated by the bimodality in the model-averaged posterior distributions.


\fi

\end{document}